\documentclass[%
 reprint,
 amsmath,amssymb,
 aps,
floatfix,
]{revtex4-2}

\usepackage{graphicx}% Include figure files
\usepackage{dcolumn}% Align table columns on decimal point
\usepackage{bm}% bold math
\usepackage{hyperref}% add hypertext capabilities
\usepackage{array}
\usepackage{booktabs}
\usepackage{float}
\usepackage{makecell}
\usepackage{subcaption}
\usepackage{tabularx,ulem}

\begin{document}

\preprint{APS/123-QED}

\title{Threshold Displacement Energies of Oxygen in YBa$_2$Cu$_3$O$_7$:\\ A Multi-Physics Analysis}% Force line breaks with \\

\author{Ashley Dickson}
%Lines break automatically or can be forced with \\
\author{Samuel T. Murphy}%
\affiliation{%
 School of Engineering, Lancaster University \\
  Lancaster, LA1 4YW, United Kingdom
}%

%\collaboration{MUSO Collaboration}
\noaffiliation

\author{Mark R. Gilbert and Duc Nguyen-Manh}
%  \homepage{http://www.Second.institution.edu/~Charlie.Author}
\affiliation{United Kingdom Atomic Energy Authority\\
Culham Campus, Abingdon, OX14 3DB, United Kingdom}
%  Second institution and/or address\\
%  This line break forced% with \\
% }%
%\author{Duc Nguyen-Manh}
%\affiliation{Materials Division, United Kingdom Atomic Energy Authority\\
%Culham Campus, Abingdon, OX14 3DB, United Kingdom}
%  Authors' institution and/or address\\
%  This line break forced with \textbackslash\textbackslash
% }%

% \collaboration{CLEO Collaboration}%\noaffiliation

\date{\today}% It is always \today, today,
             %  but any date may be explicitly specified

\begin{abstract}

Neutron bombardment of high temperature superconducting (HTS) magnets may compromise the integrity of the magnetic confinement in future fusion reactors. The amount of damage produced by a single neutron can be predicted from the threshold displacement energies (TDE) of the constituent ions in the HTS materials, such as the Rare Earth Cuperates. Therefore, in this work a Multiphysics simulation approach is adopted to determine the threshold displacement energies for oxygen in YBa$_2$Cu$_3$O$_7$. Classical molecular dynamics (MD) simulations are employed to determine statistically representative TDEs for all four oxygen sites and these results are validated using Born-Oppenheimer MD employing forces derived from Density Functional Theory (DFT). The simulations were performed at the operational temperature (25 K) and the temperature of existing neutron irradiation studies (360 K) enabling a discussion about the relevance of this data. Overall, these findings enhance our understanding of radiation-induced damage in HTS materials and provide data that can be incorporated into higher order models offering critical insights into shielding design and magnet longevity.

% \begin{description}
% \item[Usage]
% Secondary publications and information retrieval purposes.
% \item[Structure]
% You may use the \texttt{description} environment to structure your abstract;
% use the optional argument of the \verb+\item+ command to give the category of each item. 
% \end{description}
\end{abstract}

%\keywords{Suggested keywords}%Use showkeys class option if keyword
                              %display desired
\maketitle

%\tableofcontents

\section{\label{sec:level1}Introduction}

Nuclear fusion represents an attractive prospect for low-carbon energy generation\cite{roadmap}. The magnetic confinement approach to the realisation of fusion energy uses a magnetic field to confine a deuterium-tritium plasma at temperatures in excess of 100 MK \cite{ongena2016magnetic}. Scaling laws suggest that the power density in a Tokamak reactor is proportional to the fourth power of the magnetic field strength \cite{federici2024relationship}. Therefore, the development of high temperature superconducting (HTS) magnets, that can offer field strengths \(>20\) Tesla, enables construction of compact tokamaks that are more economically attractive\cite{costley2015power}.\\

The HTS magnets are based around complex tapes that typically employ a Rare Earth Barium Cuperate (REBCO)\cite{Braccini_2011} as the functional material, generally (Y,Gd)Ba$_2$Cu$_3$O$_{7-\delta}$. These materials have complex crystal structures that are susceptible to damage by the high energy neutrons produced in the fusion reaction. Therefore, understanding the nature of damage in the tapes is crucial for optimizing the required shielding, ensuring the cost-effective operation of a Tokamak using these magnets.  \\

The impact of irradiation on REBCO tapes varies, with the nature of the damage being intrinsically linked to the radiation source. Previous studies of neutron irradiated REBCO materials \cite{umezawa1987enhanced, sauerzopf1991neutron, sauerzopf1993fast, sauerzopf1995analysis, eisterer2009neutron, fischer2018effect, prokopec2014suitability} suggest that at low neutron fluences there is a near linear decrease of the critical temperature, $T_c$, and an increase in the critical current, $J_c$, due to the production of flux pinning regions in the crystal microstructure. There is a concomitant reduction in the anisotropy of $J_c$ attributed to homogeneous defect production from isotropic radiation exposure. At a certain irradiation limit (depending on the type of tape), the material undergoes a sharp decrease in superconducting properties, and eventual loss of superconductivity. This is attributed to the defect density becoming too high compared to the flux line spacing, such that defective regions no longer act as efficient pinning centres \cite{sauerzopf1991neutron}. Interestingly, the introduction of point defects seems to directly correspond to a sharp rise in J$_c$, whereas regions with larger defect density appear to have a more significant effect on $T_c$ \cite{chudy2012point}.\\

Ion irradiation can be used as a convenient proxy for studying neutron induced damage. Low fluences of ions (i.e. H$^+$, He$^+$, Ar$^{2+}$) on REBCO tapes result in similar $T_c$ degradation as observed under neutron irradiation \cite{xiong1988transport, meyer1992transport}, and the peak $J_c$ is also reproduced \cite{van1990critical}. It is generally assumed that the oxygen sublattice harbours the majority of the damage in REBCO, with Navacerrada \textit{et al.} \cite{navacerrada2000critical}, suggesting that the cation sublattices of Ba, Cu, and Rare-Earth (RE) ions are unperturbed by He$^+$ irradiation. They suggest oxygen atoms are displaced preferentially, with Valles \textit{et al.} \cite{valles1989ion} proposing specific loss of oxygen from Cu-O chains, forming interstitial defects in the vacant a-axis positions (as oxygen is the most weakly bound species, easily removed by thermal treatment \cite{swinnea1987crystal}). Extended thermal treatment results in complete loss of superconductivity, therefore, it is plausible that damage to the oxygen sublattice is responsible for the observed degradation in $T_c$. Iliffe \textit{et al.} \cite{iliffe2021situ} provide evidence for a highly mobile oxygen sublattice in GdBa$_2$Cu$_3$O$_7$, supporting this hypothesis. By contrast, Nicholls \textit{et al.} \cite{nicholls2022understanding}, using a combination of X-ray spectroscopy and Density Functional Theory (DFT) propose the formation of defects in the Cu-O$_2$ planes of REBCO under He$^+$ irradiation; in opposition to the general assumption that $T_c$ degradation is solely attributed to loss of oxygen from Cu-O chains. \\

Other computational investigations of defect chemistry and radiation damage in REBCO provide crucial insight into irradiation effects. A comprehensive description of the defect chemistry in YBCO by Murphy \cite{murphy2020point} indicates that the oxygen Frenkel process is the most thermodynamically favourable process in the material. Furthermore, high concentrations of oxygen defects are shown to result in the clustering of vacancies on the O1 site, in agreement with experimental work that suggests damage is found primarily in the Cu-O chains. Improving upon prior work from Chaplot \cite{chaplot1990interatomic} and Baetzold \cite{baetzold1988atomistic}, an updated empirical pair potential for YBCO was developed by Gray \textit{et al.} \cite{gray2022molecular}. Simulation of displacement events in the keV regime demonstrate amorphisation of the material, consistent with experimental observation of high-energy cascades \cite{strickland2023near}. Furthermore, the temperature effect on the number of defects is explored, demonstrating that more defects are produced at higher temperatures. Further cascade simulations of Torsello \textit{et al.} \cite{torsello2022expected} were coupled with neutronics simulations to indicate that at the fluence expected on the Toroidal Field (TF) coil magnet of the ARC fusion reactor concept, the displacements per atom (dpa) is estimated to be 0.52 in 10 years, a high value for a complex crystalline ceramic. \\

The Threshold Displacement Energy (TDE) is the fundamental basis of predicting damage production in materials, particularly in codes employing the binary collision model, such as SRIM \cite{ziegler2010srim}, where the TDE serves as the primary criterion for quantifying damage caused by ion irradiation. The TDE is also used in the classical displacements per atom (dpa) approximation to quantify atomic damage dose caused by nuclear radiation. If an incident neutron or high energy ion transfers, via collision/scattering, an energy greater than TDE, the impacted (target) atom can be displaced. Depending on how much energy is transferred, neighbouring atoms may also be displaced as the primary knock-on atom (PKA) initiates a collision cascade \cite{beland2016atomistic}. \\

Within a Molecular Dynamics (MD) simulation the TDE is highly dependent on the interaction model used \cite{fikar2009molecular} and the specific directions investigated. Therefore, to address these uncertainties, we perform classical MD simulations utilising the potential of Gray \textit{et al.} \cite{gray2022molecular} to determine a statistically representative threshold displacement energy for all of the oxygen sites in YBa$_2$Cu$_3$O$_7$. To demonstrate the reliability of these energies we perform a similar investigation using DFT, which provides a quantum mechanical description of the ions. This enables the exploration of processes that are not possible using empirical potentials, such as charge transfer. Previous studies have employed DFT for the calculation of TDEs in simple materials \cite{xiao2009threshold, rahman2021ab, olsson2016ab, lucas2005ab}, however, these studies lack a rigorous stochastic treatment of the quantity. TDE calculations are performed at both 25 K and 360 K to reflect the operating temperature of fusion magnets and the elevated temperatures experienced during neutron irradiation in the TRIGA reactor, where recent magnet tests were conducted \cite{sauerzopf1991neutron}.

\section{REBCO crystallography}
\begin{figure}[H]
    \centering
    \includegraphics[width=0.6\linewidth]{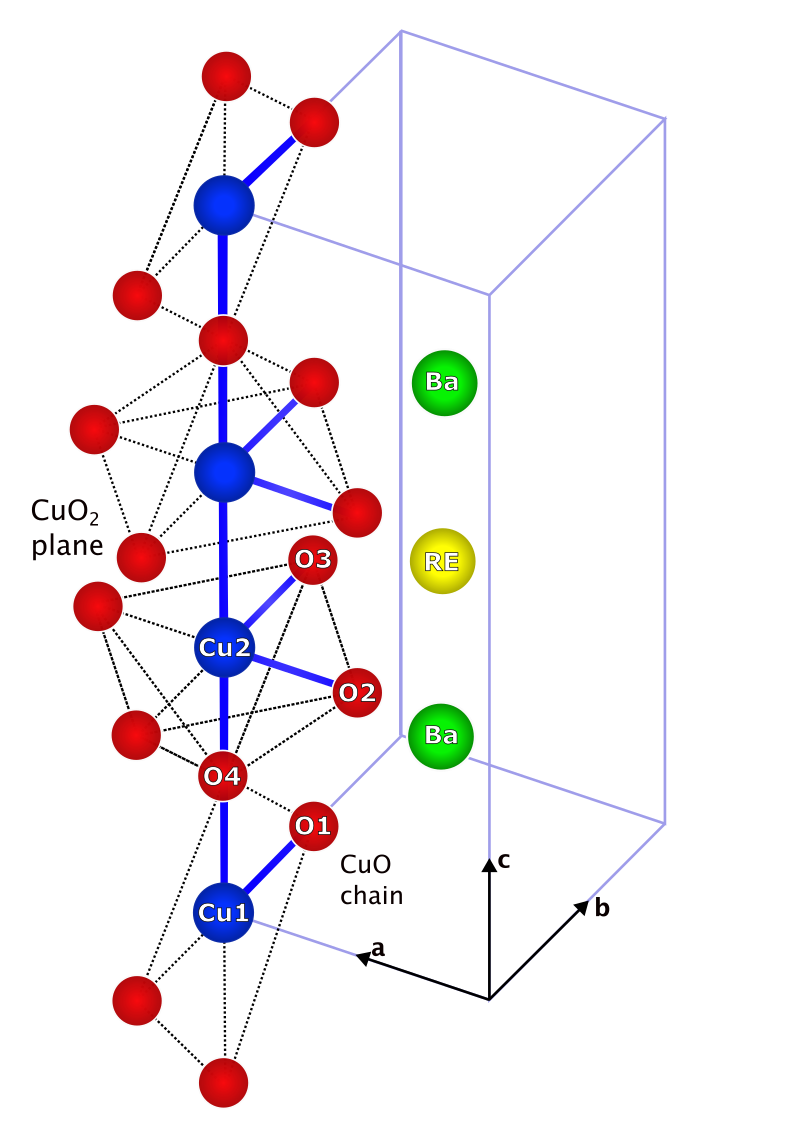}
    \caption{Crystal structure of REBCO. The central atom is exchanged with a RE element.}
    \label{REBCO}
\end{figure}

REBCO compounds display a complex and highly anisotropic crystal structure described by the $pmmm$(47) space group. Within the unit cell, two symmetrically distinct Cu sites are of note: Cu1 at the $1a$ Wyckoff site, forming Cu-O chains, and Cu2 at the $2y$ Wyckoff site, bonding with five oxygen ions in a square-based pyramidal geometry, within which the Cu-O$_2$ planes are found (illustrated in Fig. \ref{REBCO}). Decreasing the oxygen content of the crystal to sub-stoichiometric amounts (ranging from 7 down to 6) results in presumed oxygen loss from the O1 site \cite{jorgensen1990structural, kwok1988electronic} in the Cu-O chains, reducing the Cu1 coordination to 2-fold, with only bonds to O4 atoms remaining. This results in the Cu1 oxidation state shifting from +3 to +2 and a perturbation in the crystal symmetry from orthorhombic to tetragonal. As the oxygen content decreases, so does the superconducting transition temperature, $T_c$, where eventually superconducting properties are lost in their entirety \cite{benzi2004oxygen}. In this work we focus on stoichiometric YBa$_2$Cu$_3$O$_7$ (YBCO). 

\section{Methodology}\label{methods}
\subsection{DFT Simulation Details}
All DFT simulations are performed using CP2K \cite{hutter2014cp2k}, chosen due to its quasi-linear scaling with system size. TDE simulations require a box size large enough to reduce spurious finite size effects, therefore, it is critical to use a code that scales efficiently with system size. The Generalised Gradient Approximation (GGA) was employed within the Perdew-Burke-Ernzerhof (PBE) functional framework \cite{perdew1996generalized}, selected due to its robustness and computational efficiency. Atoms were represented with a mixed Gaussian and plane-wave basis (included in the \textit{Quickstep} package native to CP2K), and the norm-conserving, dual space, separable Goedecker-Teter-Hutter (GTH) pseudopotentials \cite{krack2005pseudopotentials} optimised for PBE were used (Table \ref{aa} details the explicitly considered electrons in our calculations).

\begin{table}[h!]
\begin{center}

\caption{Explicitly considered electrons in our calculations.}
\begin{tabular}{ c | c   }

Atom & Explicit electrons  \\
\hline
\hline 
Y & $4s^2, 4p^6, 5s^2, 4d^1$  \\
Ba & $5s^2, 5p^6, 5s^2$\\
Cu & $3d^{10}, 4s^1$\\
O & $2s^2, 2p^4$\\

\end{tabular}
\label{aa}
\end{center}
\end{table}

We use a 6$\times$6$\times$2 supercell (936 atoms) of YBCO, with a converged cutoff of 650 Ry. Integration is performed at the $\Gamma$ point, and atomic orbitals are represented in a Double-Zeta Valence Polarised (DZVP) basis. We employ periodic boundaries in the x, y, and z directions. Our energy minimised lattice parameters are calculated to be 3.869 {\AA}, 3.958 {\AA}, 11.854 {\AA} for the \textit{a, b,} and \textit{c} directions, respectively. These show good agreement with the experimentally determined values of 3.827{\AA}, 3.893{\AA}, and 11.699{\AA} \cite{calestani1987crystal}. Further validation of the model can be found in Appendix \ref{validation}. \\

\subsection{Defining the Threshold Displacement Energy}
Critical to the accurate determination of the dpa in irradiated materials is a robust definition of the TDE. Nordlund \textit{et al.} \cite{nordlund2006molecular} provide good arguments to suggest that the averaged lower bound of the TDE ($E_{d, ave}^{av}$), and the production probability ($E_d^{pp}$) are the most valid definitions for use in damage prediction (for a description of both quantities refer to \cite{nordlund2006molecular}). $E_d^{pp}$ has the advantage of including in-cascade annealing for higher PKA energies. Indeed, this is valid for YBCO as we find that the displacement function is non-monotonous, and previous simulation work demonstrates in-cascade annealing of damage does occur \cite{gray2022molecular}. However, for application in damage prediction codes, it is assumed that the TDE corresponds to a situation where displacement of the PKA correlates exactly to formation of a Frenkel pair. This is a problematic assumption for materials resistant to amorphisation, as recovery mechanisms are available, and are not accounted for in the lower bound of the TDE for a given direction ($E_d^l(\theta, \phi)$\cite{robinson2012systematic}). However, both experimental and simulation work strongly suggests that YBCO is highly susceptible to amorphisation \cite{strickland2023near, torsello2022expected, gray2022molecular}, suppressing recombination. Torsello \textit{et al.} \cite{torsello2022expected} demonstrate that, for the ARC reactor, neutron flux incident on the magnets remains high even after travelling through the first wall, the steel vacuum vessel, and the FLiBe molten salt tank. As such, one would expect significant amorphisation of the material due to highly energetic collisions, therefore, it appears that the $E_d^l$ proves the most sensible definition for the TDE in the context of damage prediction. $E_d^l$ is defined as the minimum energy at which an atom is displaced for a given direction, and can be averaged over all directions to give (where $\sin \theta d\theta d\phi$ amounts to the solid angle subtended by each PKA vector):

\begin{equation}
E^{av}_{d, ave} = \dfrac{\int_0^{2\pi} \int_0^{\pi} E_d^l(\theta, \phi) \sin\theta d\theta d\phi}{\int_0^{2\pi} \int_0^{\pi} \sin\theta d\theta d\phi }
\label{EDAVG}
\end{equation}

Practically speaking, this is analogous to constructing a spherical Voronoi tessellation (we use SciPy \cite{virtanen2020scipy} for this) of the displacement vectors, then weighting the TDE value in that direction by the surface area of its corresponding Voronoi cell. \\

\subsection{Determination of the Threshold Displacement Energy}

In this work, we use the Empirical Potential (EP) for YBCO fitted by Gray \textit{et al.} \cite{gray2022molecular}. All simulations performed with the EP were carried out in the Large-scale Atomic/Molecular Massively Parallel Simulator (LAMMPS) software package \cite{LAMMPS}. We initially determine the length of the thermal spike in YBCO to be 400 fs (see appendix \ref{thermspike}), and take this as the time at which the simulation is checked for defects. After conclusion of the thermal spike, any recombination that occurs should be attributed to thermally activated recombination processes, therefore, these are considered separate to the cascade event. If the system is checked for defects during the time within which these processes occur, the TDE is artificially inflated, and any predicted dpa values will be decreased. \\

We perform our TDE simulations in the same size supercell as our DFT model, applying the same methodology (unless stated otherwise) for both the DFT and classical MD simulations. The cells are equilibrated in an NPT ensemble for 1000 fs, at both 25 and 360 K. A PKA is selected near the centre of the cell, and a boundary Langevin thermostat is employed for the outer 0.5 ${\AA}$ of the supercell, with a friction coefficient of 100 fs$^{-1}$. The boundary thermostat is employed to prevent the thermal shock front interacting with the periodic image of itself. The inner region undergoes NVE MD. The PKA is given an impulse and the simulation is allowed to evolve for 400 fs. The cell is then quenched by an energy minimisation scheme, then checked for defects via Wiegner-Seitz analysis (implemented through the Ovito python pipeline \cite{stukowski2009visualization}). The PKA energy is increased if no defect is found (by 1 eV for the EP, 2 eV for DFT). Otherwise, the PKA energy is recorded as the TDE for that direction.\\

\begin{figure}[h!]
    \centering
    \includegraphics[width=\linewidth]{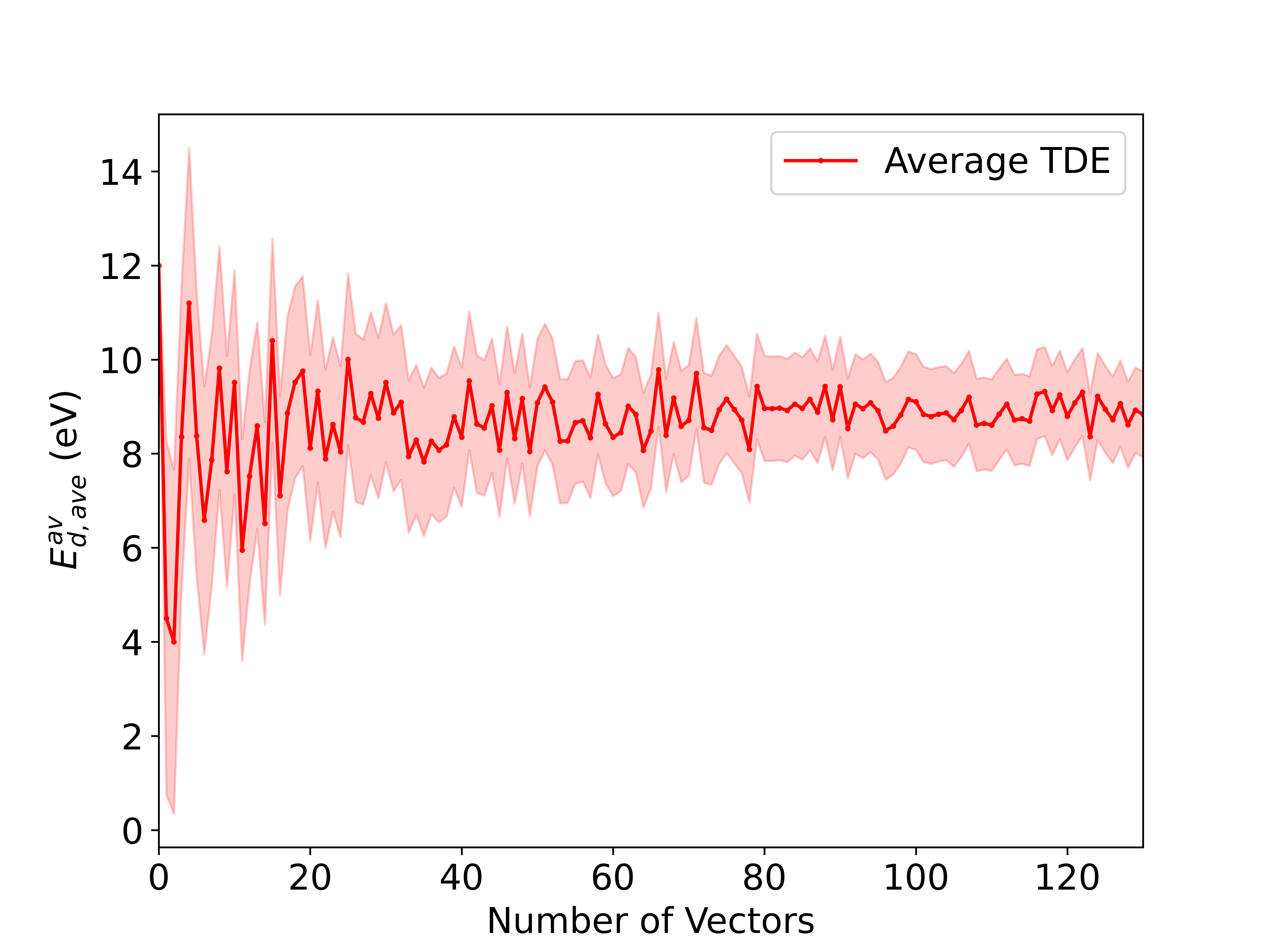}
    \caption{Effect of number of sampled directions on $E^{av}_{d, ave}$.  Simulations performed with EP for randomly selected directions at 360 K. The shaded region indicates one standard deviation.}
    \label{convgraph}
\end{figure}

Several methods exist for selecting PKA directions, such as incrementing the spherical angles in fixed steps or minimizing the energy of points distributed on a sphere \cite{robinson2012systematic}. The former method suffers from clustering points towards the poles of the unit sphere, and the latter may not fully capture the anisotropy of the TDE surface due to the even distribution of directions. Therefore, we argue that for highly anisotropic materials (such as YBCO), a random sampling approach should be employed. Vector directions for our TDE simulations are randomly selected from the unit hemisphere (with its pole directed along the \textit{a} axis), as the local environment of all oxygen atoms is symmetric about the \textit{bc} plane of YBCO. To ascertain the required number of TDE directions for a converged result, we used the EP to perform TDE simulations (using the above methodology) for $n$ ($n\in[1,130]$) random directions at 360 K (as the more noisy temperature). For each $n$, $E^{av}_{d, ave}$ was calculated, amounting to around 70,000 total simulations. Our results are shown in Fig. \ref{convgraph}. We deem the TDE to be sufficiently converged at 50 vector directions. Therefore, for all subsequent calculations of $E^{av}_{d, ave}$, the same 50 random directions were used. \\

The TDE is sensitive to the relative locations of the surrounding ions and their velocities\cite{chirita2019influence}. Therefore, to examine this we use the empirical pair potential to create unique starting configurations, obtained by modifying the seed for initializing the velocities in LAMMPS. We choose 50 different seeds for each oxygen PKA, giving 2500 total individual TDE simulations for each PKA, for each temperature. Each individual TDE simulation is repeated several times (due to the energy incrementation), resulting in 25 to 50 thousand simulations per oxygen atom per temperature (for the EP). DFT simulations sample a single initial configuration for the 50 directions. For our DFT calculations, around 200 simulations were performed at each temperature. \\

\section{Results}

\subsection{Empirical Potential TDE}
\begin{figure*}[h!]
    \centering
    \begin{subfigure}[t]{0.49\textwidth}
    \includegraphics[width= \linewidth]{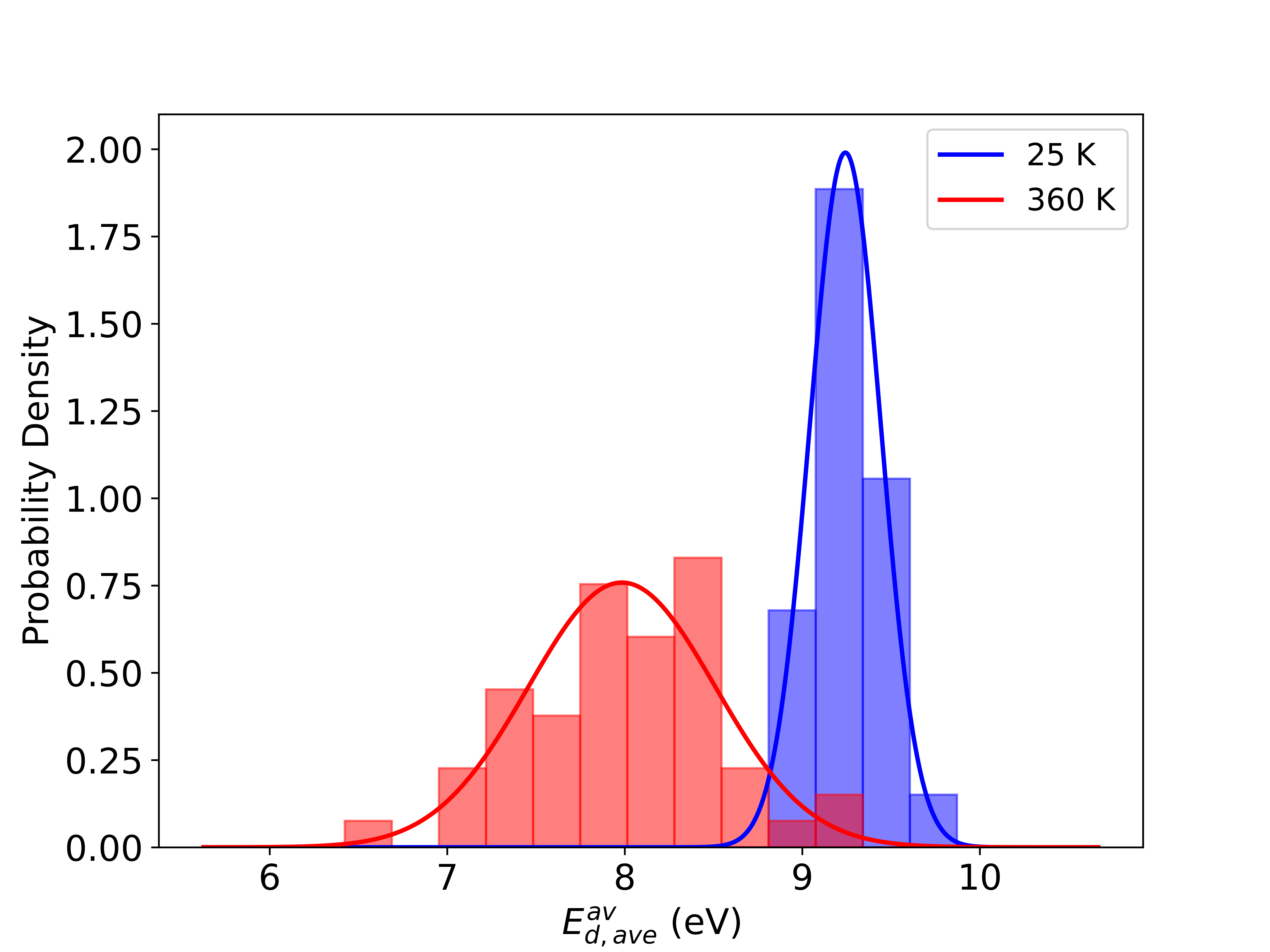}
    \caption{O1}
    \end{subfigure}
    \begin{subfigure}[t]{0.49\textwidth}
    \includegraphics[width = \linewidth]{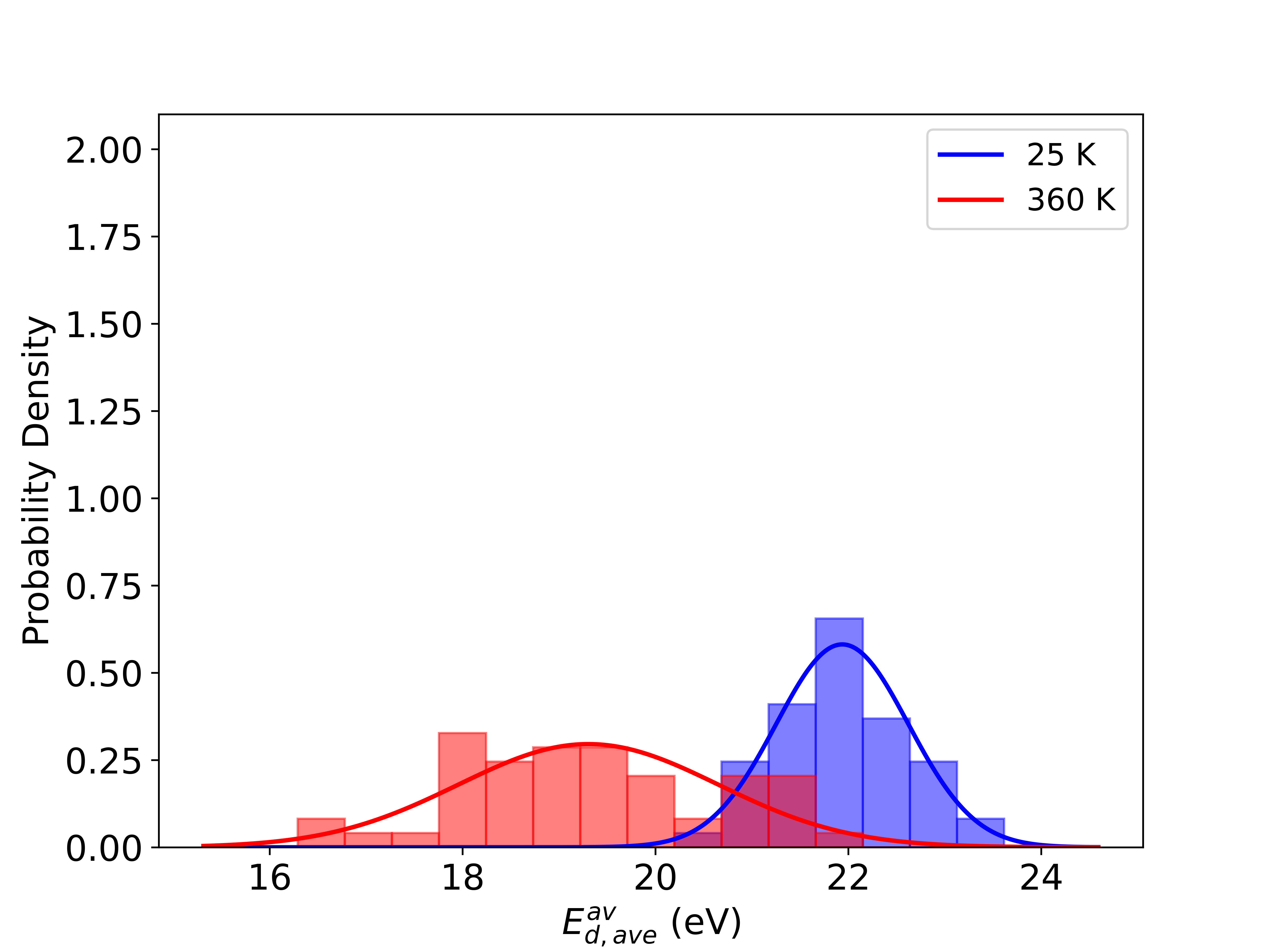}
    \caption{O2}
    \end{subfigure}
    
    \begin{subfigure}{0.49\textwidth}
    \includegraphics[width=\linewidth]{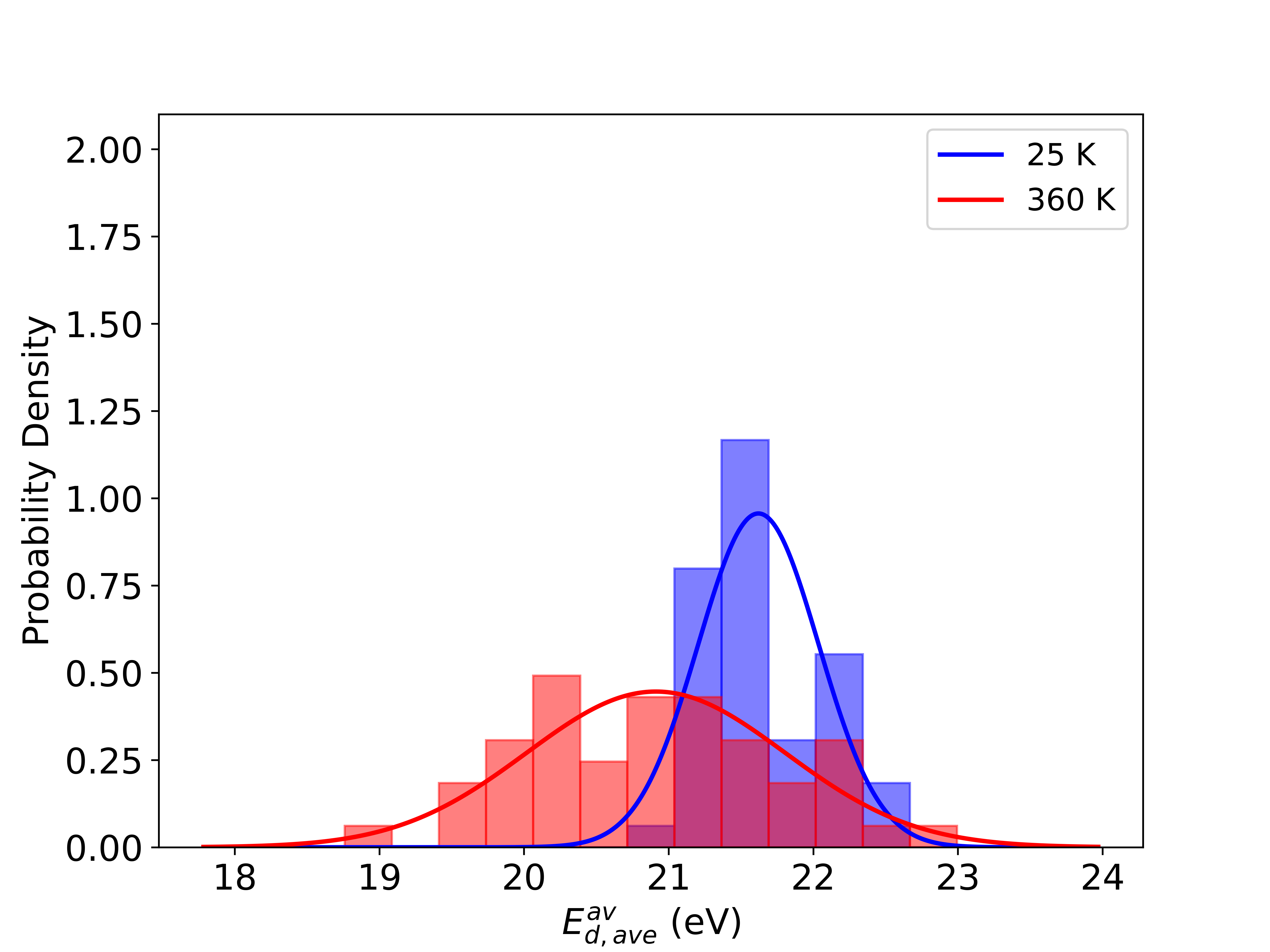}
    \caption{O3}
    \end{subfigure}
    \begin{subfigure}{0.49\textwidth}
    \includegraphics[width= \linewidth]{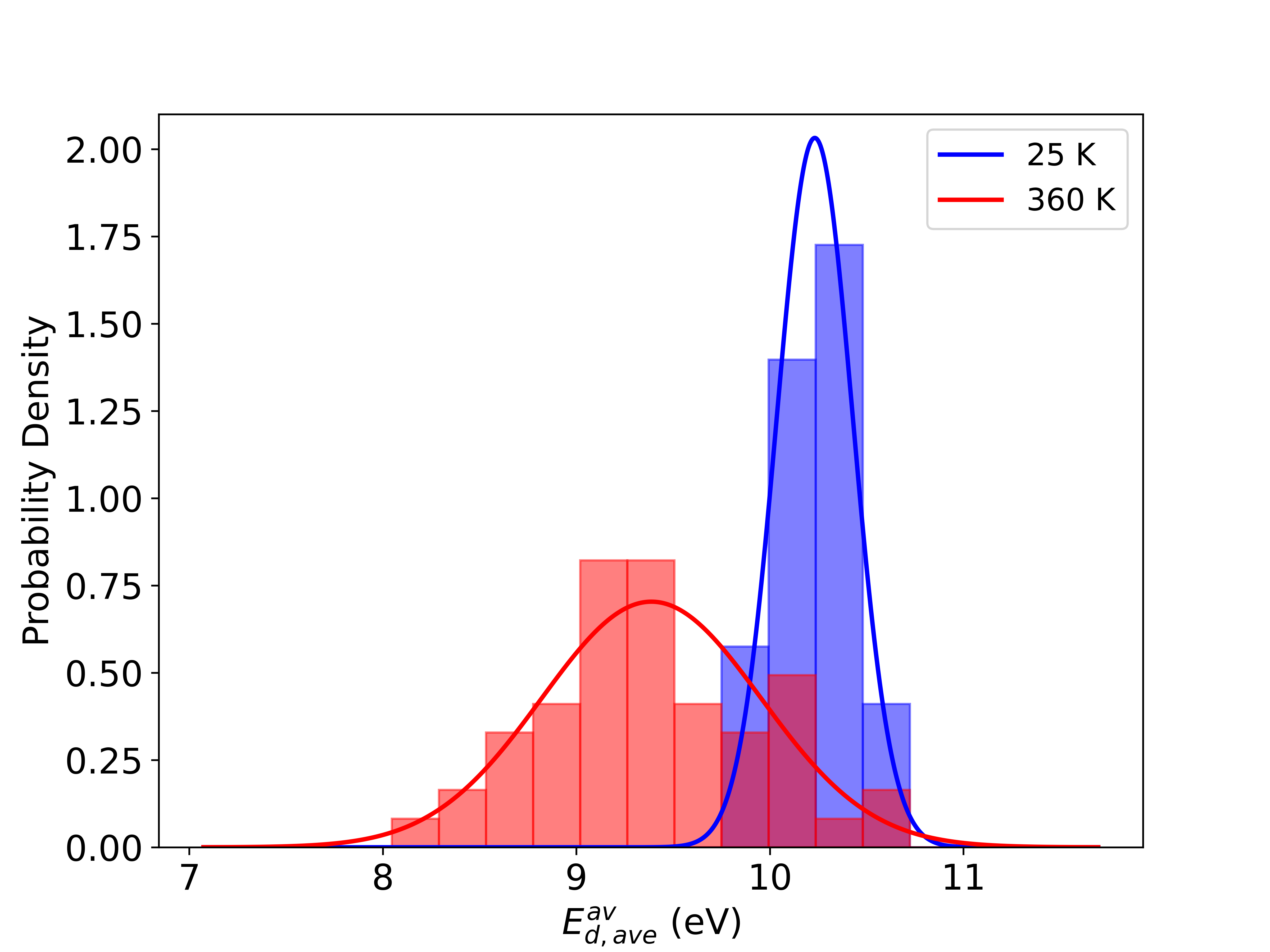}
    \caption{O4}
    \end{subfigure}
    \caption{Distributions of TDE ($E_{d,ave}^{av}$) for EP simulations from differing initial configurations. The histogram of values is overlaid with a fitted Gaussian distribution. The red distribution corresponds to simulations at 360 K, and blue corresponds to 25 K.}
    \label{EPgauss}
\end{figure*}

The TDEs determined using EP at the temperatures of 25 and 360 K, are presented in Fig. \ref{EPgauss}. The corresponding mean and variance are shown in Table \ref{meanandvar}. These results support the general presumption that oxygen atoms in the chain region are easier to displace than those in the planes. Furthermore, the energy range that is considered here agrees well with prior simulation and experimental work. In all cases, the average TDE is reduced for the higher temperature, presumably due to lattice vibrations reducing the energy barrier to atomic displacement. Observation of a decrease in TDE with increasing temperature is consistent with other work that considers the lower bound of the TDE \cite{beeler2016effect, chen2019atomistic}. The variance of the TDE appears to increase not only with temperature, but also with PKA energy. As discussed in graphene by Chirita \textit{et al.} \cite{chirita2019influence}, the momenta of neighbouring atoms to the PKA have a direct effect on the ease of displacement. For higher temperatures, these effects are more pronounced, either more drastically aiding or deterring displacement. Secondary to this, perturbations in the crystal lattice can lead to alterations in $E_d^l(\theta, \phi)$, even at zero temperature \cite{chirita2019influence}. For higher PKA energies, on average more collisions occur, where each collision has a range of outcomes based upon the velocity of the colliding atoms. The result of this is a broadening of the distribution. In addition, for higher PKA energies, channelling effects may also increase the variance in the TDE, although it is unlikely the PKA range is sufficiently large here to activate such effects \cite{nordlund2016large}.  \\

\begin{table}[h!]
\centering
\caption{Mean ($\mu$) and variance ($\sigma$) of $E_{d,ave}^{av}$ for different oxygen PKAs at 360 and 25 K (using EP).}
\begin{tabularx}{\linewidth}{
    >{\hsize=1\hsize \centering\arraybackslash}X|
    >{\hsize=1\hsize \centering\arraybackslash}X
    >{\hsize=1\hsize \centering\arraybackslash}X|
    >{\hsize=1\hsize \centering\arraybackslash}X
    >{\hsize=1\hsize \centering\arraybackslash}X
  }

    \toprule
    \multicolumn{1}{c|}{} & \multicolumn{2}{c|}{25 K} & \multicolumn{2}{c}{360 K} \\
    \hline
    PKA & $\mu$ (eV) & $\sigma$ (eV) & $\mu$ (eV)& $\sigma$ (eV) \\
    \hline 
    \hline
    O1 & 9.24 & 0.04 & 7.99 & 0.28 \\
    O2 & 21.94 & 0.47 & 19.30 & 1.82\\
    O3 & 21.62 & 0.17 & 20.91 & 0.80 \\
    O4 & 10.23 & 0.04 & 9.39 & 0.33 \\
    \bottomrule

\end{tabularx}
\label{meanandvar}
\end{table}

\begin{figure}[H]
    \centering
    \includegraphics[width=0.49\linewidth]{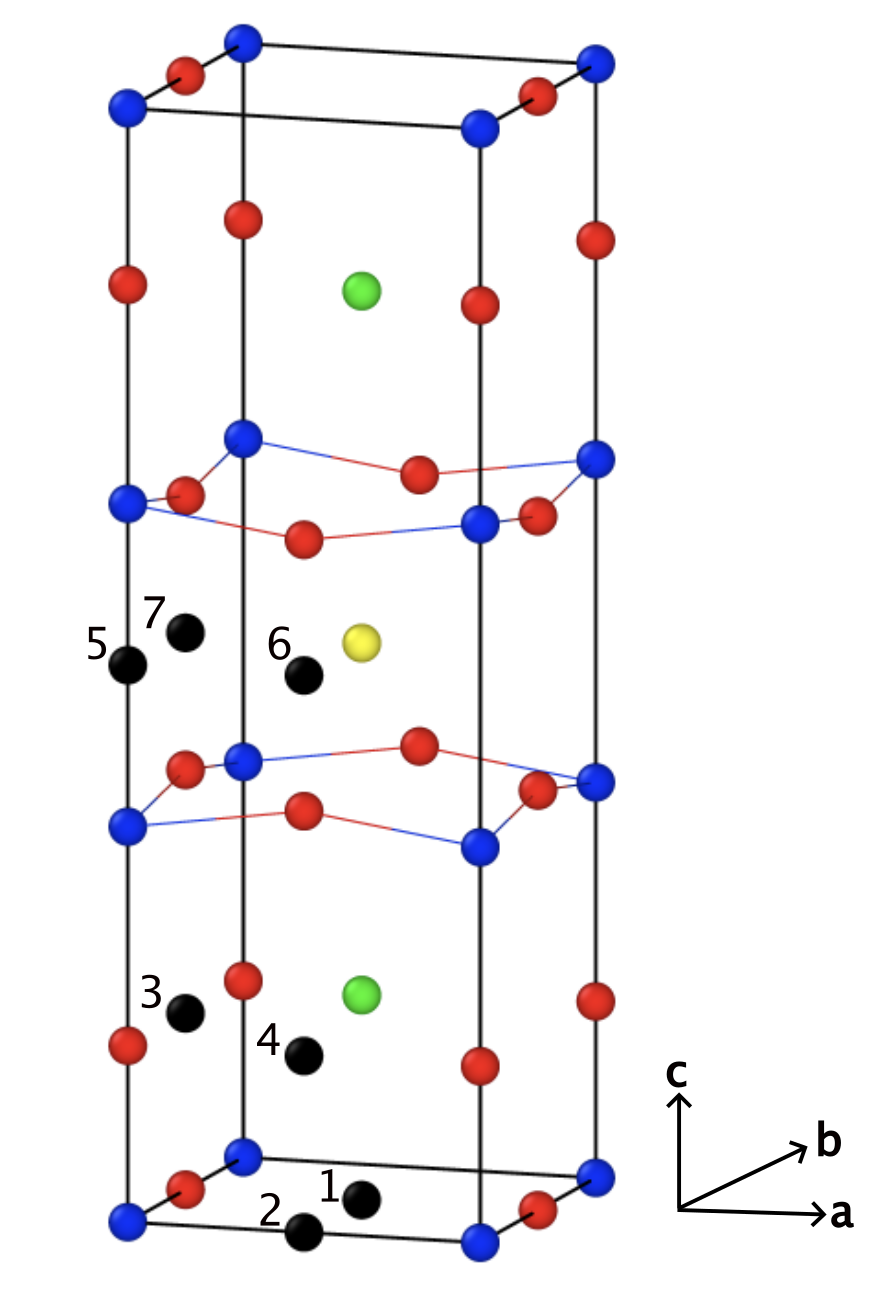}
    \caption{Labelled interstitial positions (in black) in YBCO. Cu is blue, Ba is green, Y is yellow and O is red.}
    \label{intersititals}
\end{figure}

\begin{figure*}
    \centering
    \begin{subfigure}[t]{0.49\textwidth}
    \includegraphics[width= \linewidth]{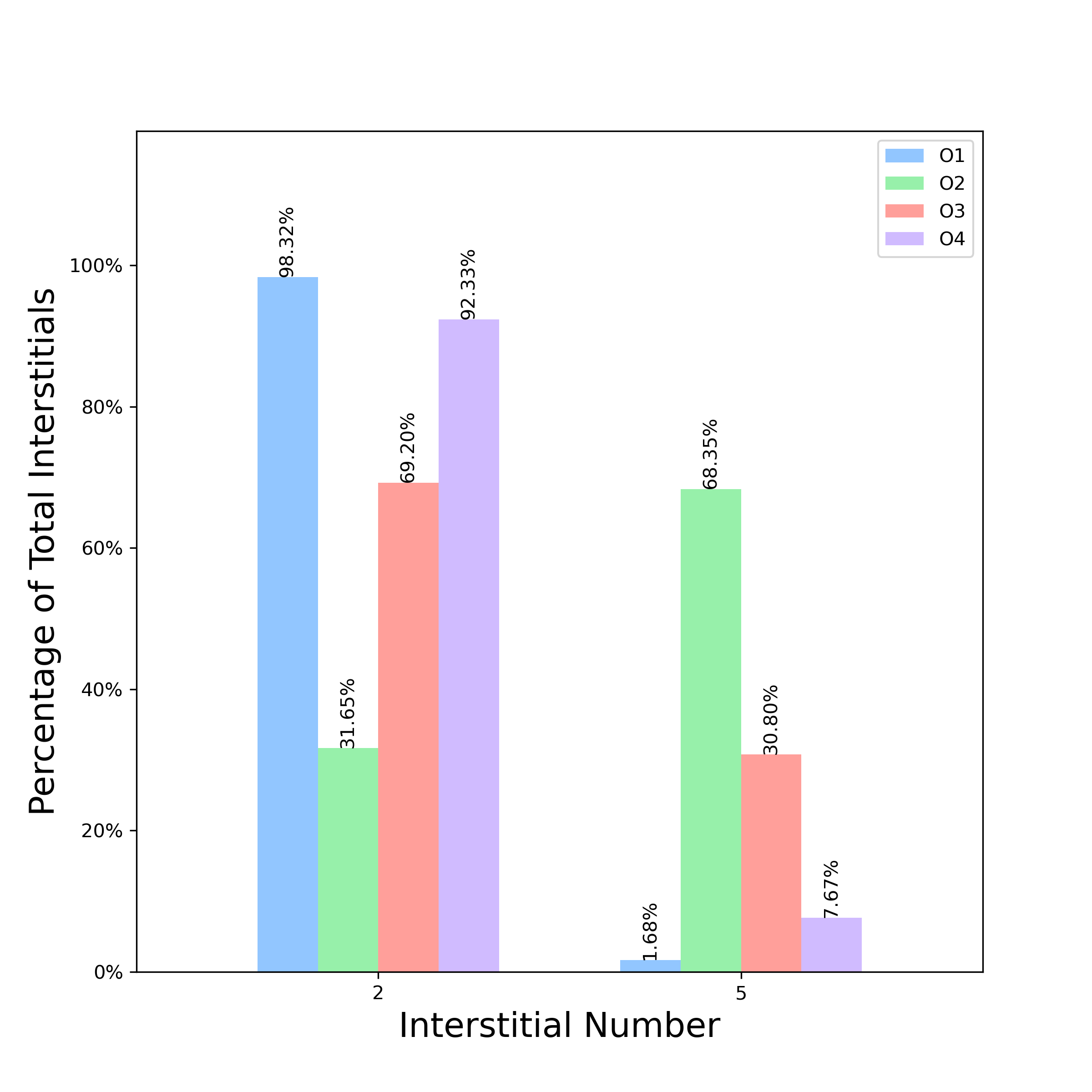}
    \caption{25 K}
    \end{subfigure}
    \begin{subfigure}[t]{0.49\textwidth}
    \includegraphics[width = \linewidth]{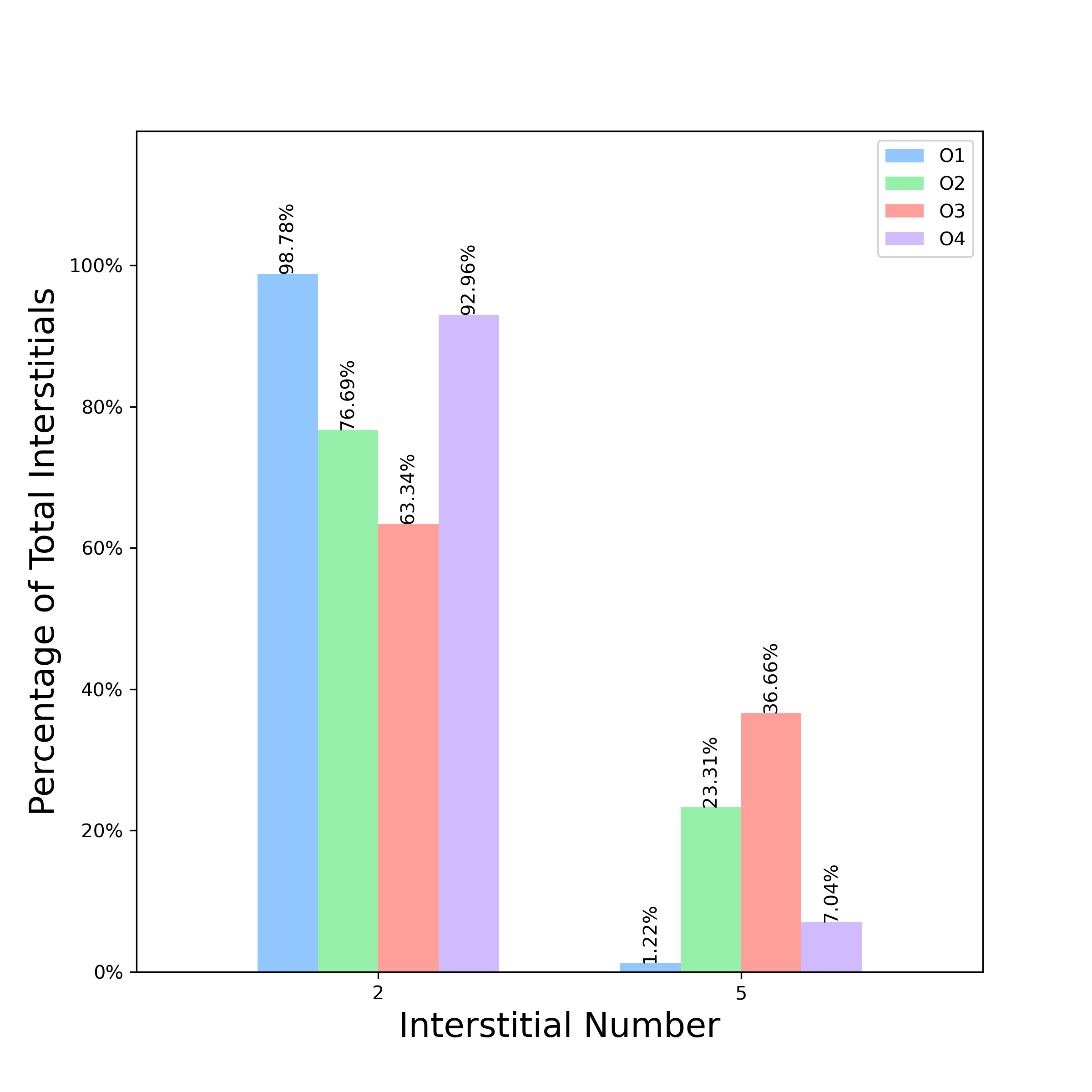}
    \caption{360 K}
    \end{subfigure}
    
    \caption{Percentage of different interstitial types observed for each oxygen PKA at 25 and 360 K. Any interstitials with total occupancy below 1.5 \% are not included.}
    \label{defbar}
\end{figure*}

The different interstitial types in YBCO are described in Fig. 
 \ref{intersititals}, and the percentages of different interstitial defects generated from our simulations are shown in Fig. \ref{defbar}. It is clear that in most cases, interstitial formation is dominated by the 2 site, closely followed by the 5 site. The concentration of interstitials in the 5 site is increased for the plane PKAs (O2 and O3). This is to be expected as these PKAs have a smaller spatial separation to the 5 site than the chain oxygen. This same reasoning explains why the concentration of interstitial 2 defects is higher for the chain oxygen PKAs than those in the plane. Furthermore, the results we see here are to be expected by reference to the interstitial energies in 
 \cite{gray2022molecular}, as the 2 site has the lowest formation energy, closely followed by the 5 site. An interesting observation from Fig. \ref{defbar} is that at 25 K for the O2 PKA, interstitial formation is dominated by the 5 position, whereas at 360~K we observe predominant formation of interstitial defects in the 2 position. We speculate that this is due to reduced oxygen mobility at 25~K, favouring the nearby 5 site. For higher temperatures the increased oxygen mobility may facilitate extended displacement or diffusion of the PKA to the highly stable 2 site.  \\

\subsection{DFT TDE}
\begin{figure}[h!]
   \centering
\begin{subfigure}{0.49\textwidth}
    \includegraphics[width=1\linewidth]{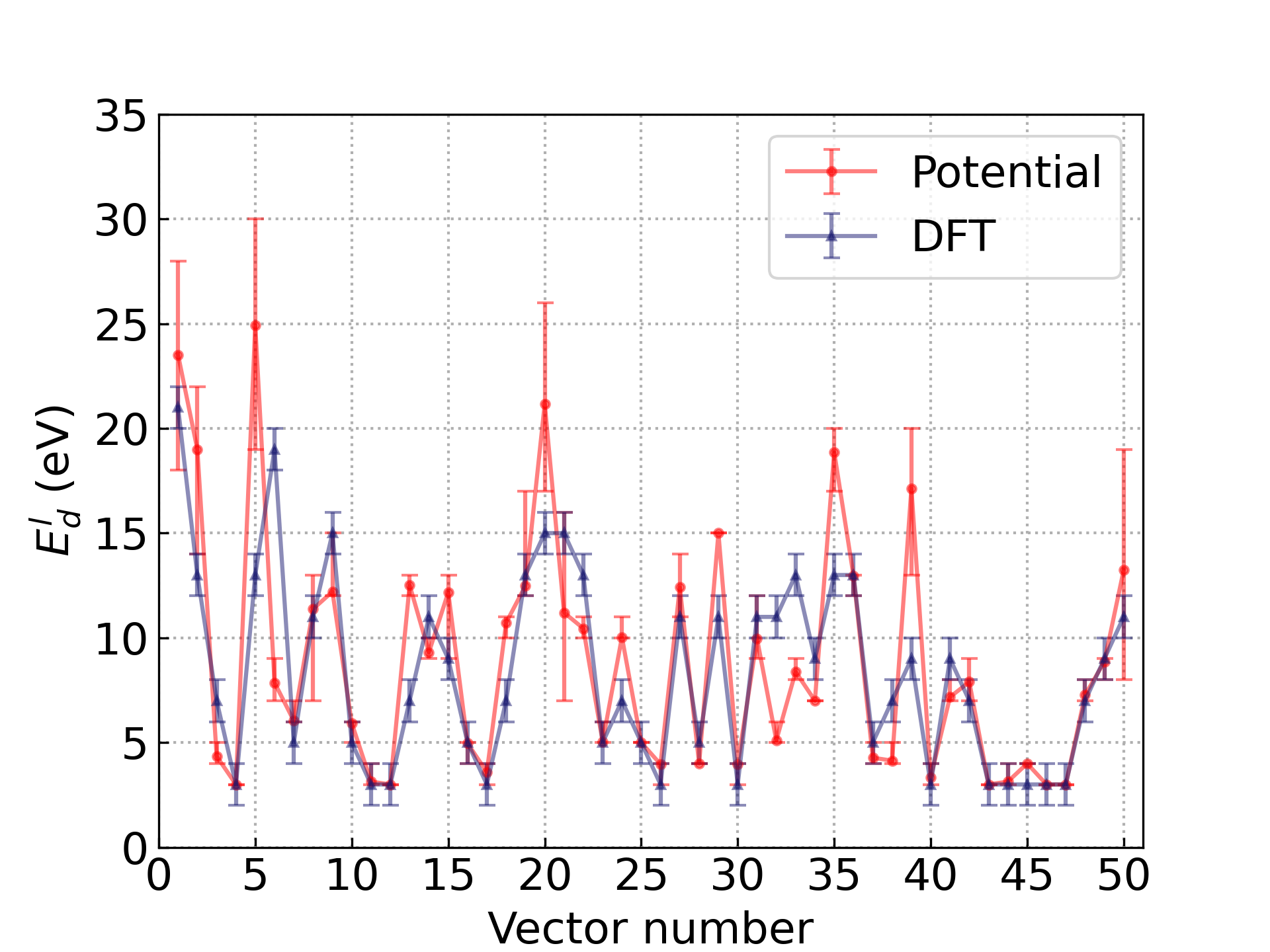}
    \caption{25 K}
\end{subfigure}

\begin{subfigure}{0.49\textwidth}
    \includegraphics[width=1\linewidth]{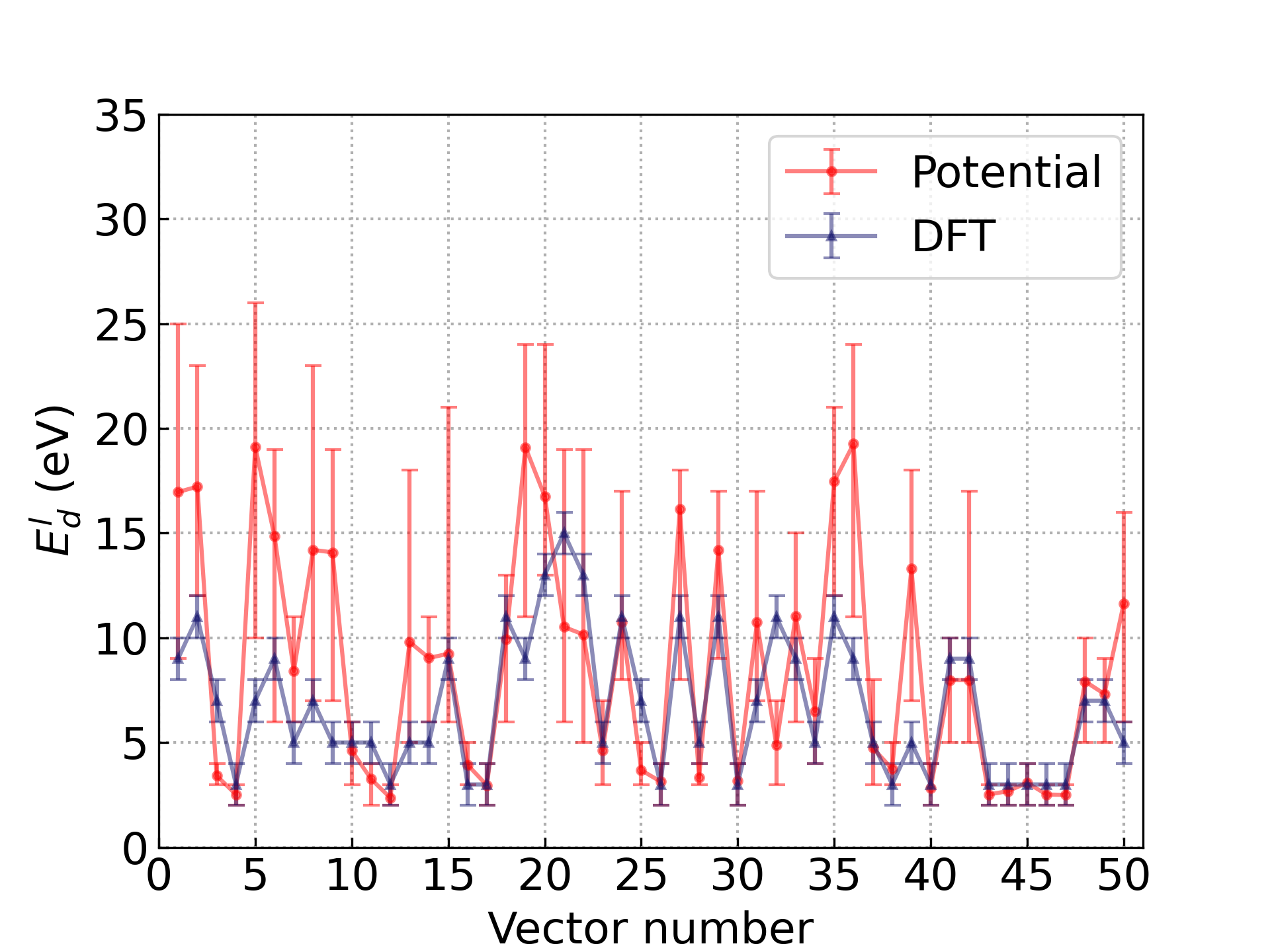}
    \caption{360 K}
\end{subfigure}
    
    \caption{Comparison of TDE ($E_d^l(\theta, \phi)$) values from EP and DFT simulations at 25 and 360 K. Vector numbers correspond to one vector in a list of 50 random vectors. Vectors are kept the same for all simulations, so direct comparisons can be made. The error bars for the DFT results represent the PKA discretisation error, and the error bars for the potential represent the range of values obtained due to different starting configurations.}
    \label{DFTTDE25}
\end{figure}

In order to validate the results we have obtained using the EP, the TDE of the O1 atom is compared to DFT. Fig. \ref{DFTTDE25} compares the TDE values ($E_d^l(\theta, \phi)$) from our DFT simulations with those derived from EP simulations. The values are compared across a standardised list of 50 random vectors, and the EP results are presented as the range of values obtained from differing initial configurations. Displacement vectors that correspond to direct or glancing collisions with another atom show the highest variance due to initial configuration. This effect is explained similarly to how the distribution of $E_{d,ave}^{av}$ arises. It is therefore unsurprising that atomic displacements directly towards a vacant interstitial position result in a $E_d^l(\theta, \phi)$ with little dependence on initial velocities of the neighbouring ions. \\

For simulations conducted at 25 K, we observe good agreement, with an $E_{d,ave}^{av}$ of 9.24 eV for the O1 PKA using the EP, and 7.88 eV using DFT. At 360 K we observe similar trends, with a higher $E_{d,ave}^{av}$ for the EP (7.98 eV) than DFT (7.25 eV). The reduction of the TDE with higher temperature is expected when considering the lower bound of the TDE. Our results (in both cases) show excellent agreement with prior studies on the TDE of YBCO. Gray \textit{et al.} \cite{gray2022molecular} report a range of values for the symmetrically distinct oxygen atoms consistent with those we present here. Previous computational work indicates that $E_d^l(\theta, \phi)$ can be as low as 1.5 eV \cite{cui1992preferential}, with certain directions corresponding to higher values \cite{cui1994simulations}. Our simulations clearly show a number of instances where $E_d^l(\theta, \phi)$ is as low as 2 eV (Fig.~\ref{DFTTDE25}). Experimental work on TDEs in YBCO is limited, however, electron irradiation suggests a displacement energy for the O1 atom of 18-20 eV at $\sim$80 K \cite{giapintzakis1994testing, mitchell1988electron, basu1991electron}. Similar work from Legris \textit{et al.} suggests a TDE of as low as 10 eV for the O1 atom\cite{legris1993effects}. These irradiations are performed near the $c$-axis, therefore, even when accounting for beam spreading \cite{nordlund2006molecular}, there will be a bias such that these results are not directly comparable with our DFT results. Nevertheless, it is encouraging that our TDEs are close to experimental values. We also note that thermal recombination results in an increased TDE from experiments, perhaps accounting for the observed differences.  \\

% A further note relates to the equilibration time of the two simulation methods. We use a constant time for all simulations (400 fs), however, the equilibation time for the two methods differs. The EP equilibrates far quicker than the DFT (appendix \ref{thermspike}), therefore, it is plausible that recombination processes can occur in this short time window. We have performed simulations with the EP showing rapid, spontaneous annihilation of defects with no temperature dependence, on timescales $<$ 100 fs. This may ``artificially'' elevate the TDE we obtain from the EP, however, for validation purposes, it is important that simulations are kept the same between the DFT and EP.\\

\begin{figure}[H]
    \centering
    \includegraphics[width=1\linewidth]{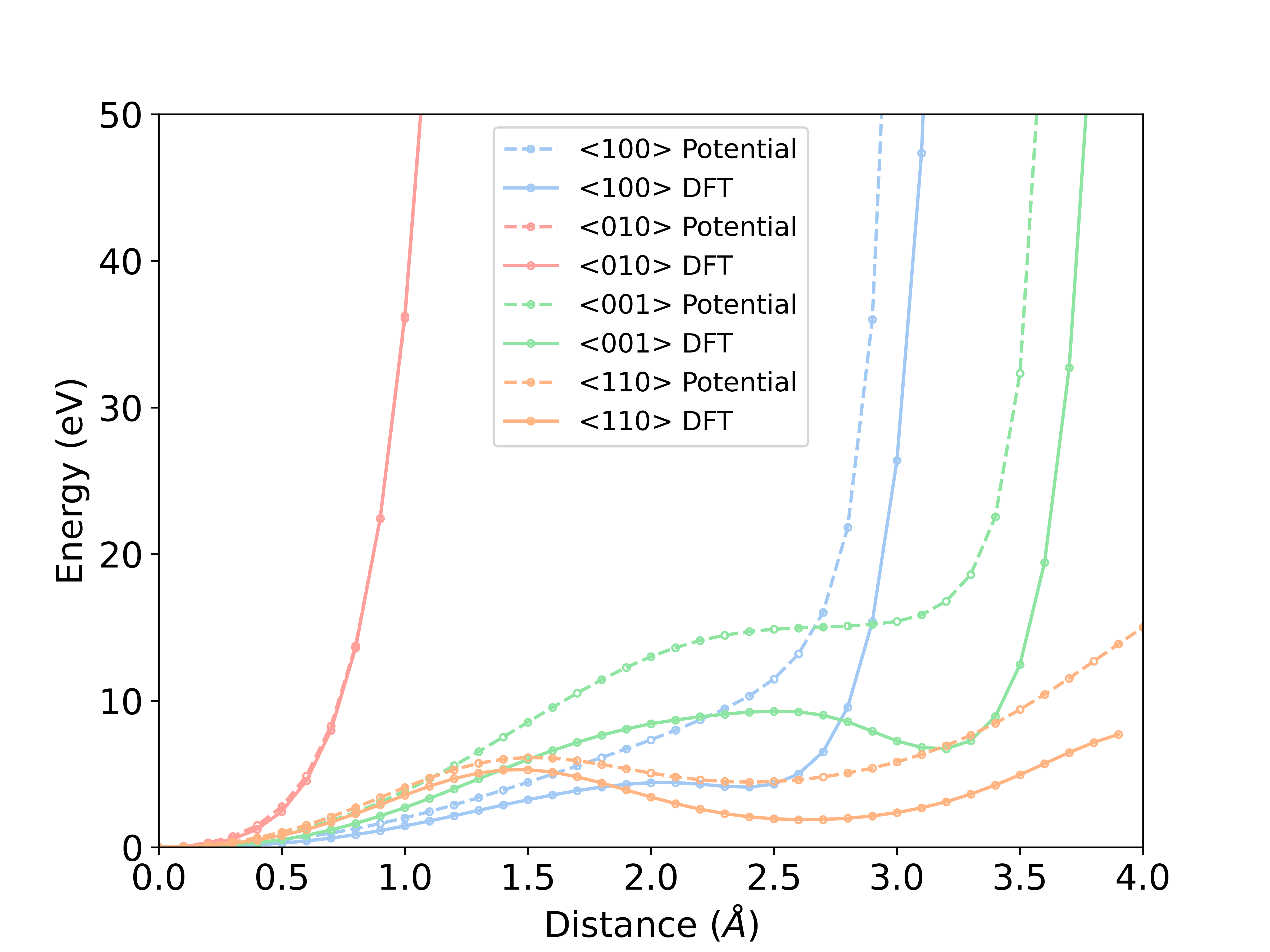}
    \caption{QSD calculations for displacement of an O1 atom along $<100>$, $<010>$, $<001>$, and $<110>$ directions. Performed for both the DFT and EP models. }
    \label{QSD}
\end{figure}

The treatment of close atomic approaches is a significant point of difference between DFT and the EP. The EP is splined to a Ziegler-Biersack-Littmark (ZBL) \cite{ziegler2010srim} potential at short distances, which is specifically designed to represent short range interactions. No such interaction model is included for the DFT model, where one must consider the cutoff radius of the pseudopotential, below which the wavefunction becomes entirely nonphysical. We find a minimum approach distance of $\sim$1.1 {\AA} between two oxygen atoms in our DFT simulations. The non-local cutoff distance for oxygen with the GTH pseudopotentials is 0.127 {\AA}, therefore the interactions in our simulations should remain valid. It is worth noting that beyond these cutoff radii the pseudo-wavefunction and real wavefunction do not strictly coincide, rather they approach each other at a rate proportional to the cutoff radius. As such, ideally atoms approach each other at a distance far higher than this cutoff. To support our argument that the pseudopotentials are sufficient, we have performed Quasi-Static Drag (QSD) calculations (Fig. \ref{QSD}). This analysis provides an understanding as to how different interaction potentials respond to the step-wise displacement of a target atom. We perform QSD simulations by displacement of the O1 atom in YBCO across the three principle crystallographic axes, and along the $<110>$ direction (towards the vacant interstitial 2 position). In general, for the distances considered here, the pseudopotential agrees very well with the ZBL potential. Therefore, we deem the DFT short range interactions to be sufficient for these low energy cascades.  This conclusion is based on an analysis of the energetic response to close atomic approaches, particularly in comparison to the maximum PKA energy used in our simulations ($\approx$ 30 eV). Notably, the DFT model captures several local minima along the chosen crystallographic axes, features that are not represented by the EP. This further highlights the importance of performing TDE simulations using a more explicit interaction model.\\

\begin{figure}[h!]
\centering 
\includegraphics[width = \linewidth]{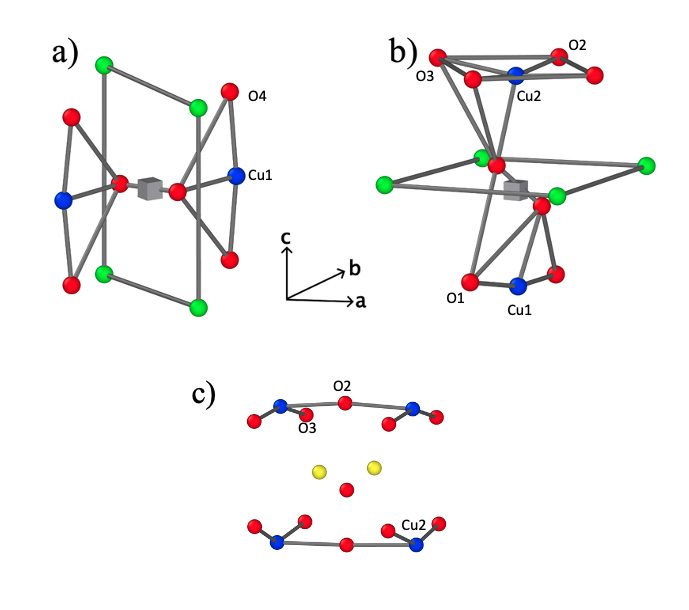}
\caption{Morphology of different defect structures observed from DFT TDE simulations in YBCO. Cu atoms are blue, Y is yellow, Ba is green and O is red. The panels correspond to the following defects: a) O$_2$ peroxide dumbbell, sat in ab plane next to O1 vacancy in CuO chain b) O$_2$ peroxide dumbbell next to O4 vacancy. Both of these defects have an O-O bond length of 1.45 ${{\AA}}$. c) distorted interstitial 6 defect.}
\label{defectmorph}
\end{figure}

To further probe the differences between the models, we have also analysed the structure of the defects that arise from our simulations. The proportion of these different defects is given in Table \ref{defects}. Both the DFT and EP results predict the same predominance of oxygen interstitial defects in position 2 (symmetrically distinct interstitial positions are shown in Fig. \ref{intersititals}). In our DFT simulations, we observe formation of an oxygen interstitial in position 6 (see Fig. \ref{defectmorph}), which is not observed at any point in our EP simulations. The defect formation energies from \cite{gray2022molecular} show the interstitial 5 is markedly more favourable for the EP than the interstitial 6 (defect energies of -5.51 eV and -1.15 eV, respectively). Prior DFT work from Murphy \cite{murphy2020point} emphasizes that the interstitial 6 is not stable, and collapses down to the interstitial 5 on relaxation. It is plausible that the defect is stabilised by an O1 vacancy in the same unit cell, which we find to have a separation to the interstitial of 6 {\AA}. This vacancy may accommodate distortion of the unit cell such that the defect is stable, explaining the significant distortion observed in Fig. \ref{defectmorph}. Calculation of the separation distances of observed Frenkel pairs for our DFT simulations demonstrates the vast majority of defects formed correspond to direct displacement of the chain O1 atom into the vacant a-axis position (as suggested in \cite{valles1989ion}). A critical difference between the models is the formation of more complex defect structures in the DFT simulations, namely peroxide split interstitials (classified as other in table \ref{intersititals}). These interstitials have not been classified in detail due to the number of different structures. The morphology of two representative species is shown in Fig. \ref{defectmorph}.  We identified these species as peroxides by analysis of the length of the O-O bond, which ranges from 1.45-1.54~{{\AA}}, depending on orientation. Local distortion from neighbouring cations is likely responsible for the range of bond lengths observed. \\

\begin{table}[h!]
    \centering
    \caption{Proportion of different defect types from DFT simulation.}
    \begin{tabular}{c|c|c}
       Interstitial type  & \makecell{Occurrence (\%)\\25 K} & \makecell{Occurrence(\%)\\360 K}  \\
       \hline\hline
        2 & 82 & 74 \\
        
        5 & 0 & 6 \\
        
        6 & 2 & 0 \\

        Other & 16 & 20

    \end{tabular}
    \label{defects}
\end{table}

It is known that peroxide defects occur in metal oxides, with validation from both experiment \cite{middleburgh2014peroxide} and simulation \cite{murphy2013anisotropic, erhart2005first, youssef2012intrinsic}. One would expect that the availability of these structures will drive the TDE to lower values compared to the EP. Furthermore, these structures may also act as stable intermediates for the extended diffusion of oxygen, potentially explaining the highly mobile oxygen sublattice suggested by Iliffe \textit{et al.} \cite{iliffe2021situ}.  \\

The results in Table \ref{defects} suggest that higher temperatures facilitate extended migration of the oxygen PKA, leading to increased numbers of interstitial 5 defects. This is not prevalent for the EP. A further point of difference between the DFT and EP simulations is the formation of multiple interstitials in a single TDE simulation. Several DFT cascades at 360 K (and one instance at 25 K) show the formation of a double interstitial 2 defect, with clustered O1 vacancies. Murphy \cite{murphy2020point} suggests that this clustering of O1 vacancies is indeed favourable. We find that moving the O1 atom of an entire layer of CuO chains into the vacant interstitial 2 site has a formation energy of 0.32 eV, therefore it is plausible that these large scale rearrangements may occur within collision cascades. This behaviour is not identified by the EP, suggesting that larger-scale collision cascades may lack detail regarding defect clustering -- a property with significant implications for the material's superconductivity.\\

\section{Conclusion}

Our DFT results show TDEs that are slightly lower than those presented in previous work, but largely agree well with prior observations. Higher temperatures act to decrease the TDE, in agreement with work on other materials. Despite the reduced TDE at higher temperatures, simulation and experimental data demonstrate that the oxygen sublattice is highly mobile in YBCO; therefore, it is likely that longer time-scale defect recovery processes will significantly alter any damage predictions. \\

The nature of the defects resulting from different PKAs has also been explored. The results for the O1 atom show excellent agreement between the DFT and EP models, with the predominant formation of interstitial oxygen defects at site 2. Higher temperatures for the DFT simulations aid the displacement of the PKA and encourage the formation of interstitial 5 defects. Additionally, our DFT simulations demonstrate a variety of more exotic split interstitial peroxide structures that are not observed in the EP simulations. To our knowledge, these structures have not yet been observed in YBCO. Despite the small differences in the resultant defect chemistry of the damaged lattice, the DFT and EP simulations show excellent agreement. This suggests that the empirical potential can indeed be used to obtain the TDEs of the oxygen sites in YBCO with reasonable accuracy. \\

To conclude, we have determined the TDE of O atoms in YBCO using an existing EP, validated by an extensive DFT study. To our knowledge, we provide the most statistically informed and rigorous determination for TDE values in YBCO thus far. For use in damage prediction codes and/or analytical damage models, we recommend averaged TDE values of 16.8 and 15.3 eV (the combined average of all of our EP TDE values), for temperatures of 25 and 360 K, respectively. We also determine the distribution of TDE values due to initial configuration, finding that extensive variability exists in the data, even for low temperatures. This variability is further enhanced by temperature and PKA energy. Moreover, we demonstrate that large-scale, statistically relevant TDE simulations can be performed in a DFT framework. \\

\begin{acknowledgments}

This project was funded by Lancaster University and agreement 2077239. DNM and AD would like to thank EUROfusion for providing HPC facilities on the Marconi and Leonardo machines. MRG and DNM also acknowledge funding by the EPSRC Energy Programme (EP/W006839/1). Via our membership of the UK's HEC Materials Chemistry Consortium, which is funded by EPSRC (EP/R029431 and EP/X035859), this work used ARCHER2 UK National Supercomputing Service (http://www.archer2.ac.uk).

\end{acknowledgments}

\newpage

\appendix

\section{Validating the DFT model}\label{validation}

The DFT model was validated by comparing thermal expansion, defect formation energies, band structure, and density of states data.

\subsection{Thermal expansion}
The thermal expansion was calculated by measuring the averaged lattice constants of YBCO at temperatures from 10-310 K, in increments of 20 K. A 3$\times$3$\times$1 supercell of YBCO was equilibrated for 1000 fs (timestep of 1 fs) with a 2$\times$2$\times$2 \textit{k}-point grid using Langevin dynamics. A velocity rescaling thermostat was used in this time to speed up equilibration. A Nose-Hover thermostat (time constant of 100 fs) was then employed and the ensemble was switched to NPT with anisotropic volume control. The simulation evolved under these conditions for 500 fs, then the averaged lattice parameters for different temperatures were obtained. The results are shown in Fig. \ref{thermexp}:

\begin{figure}[h!]
\centering
\includegraphics[width=\linewidth]{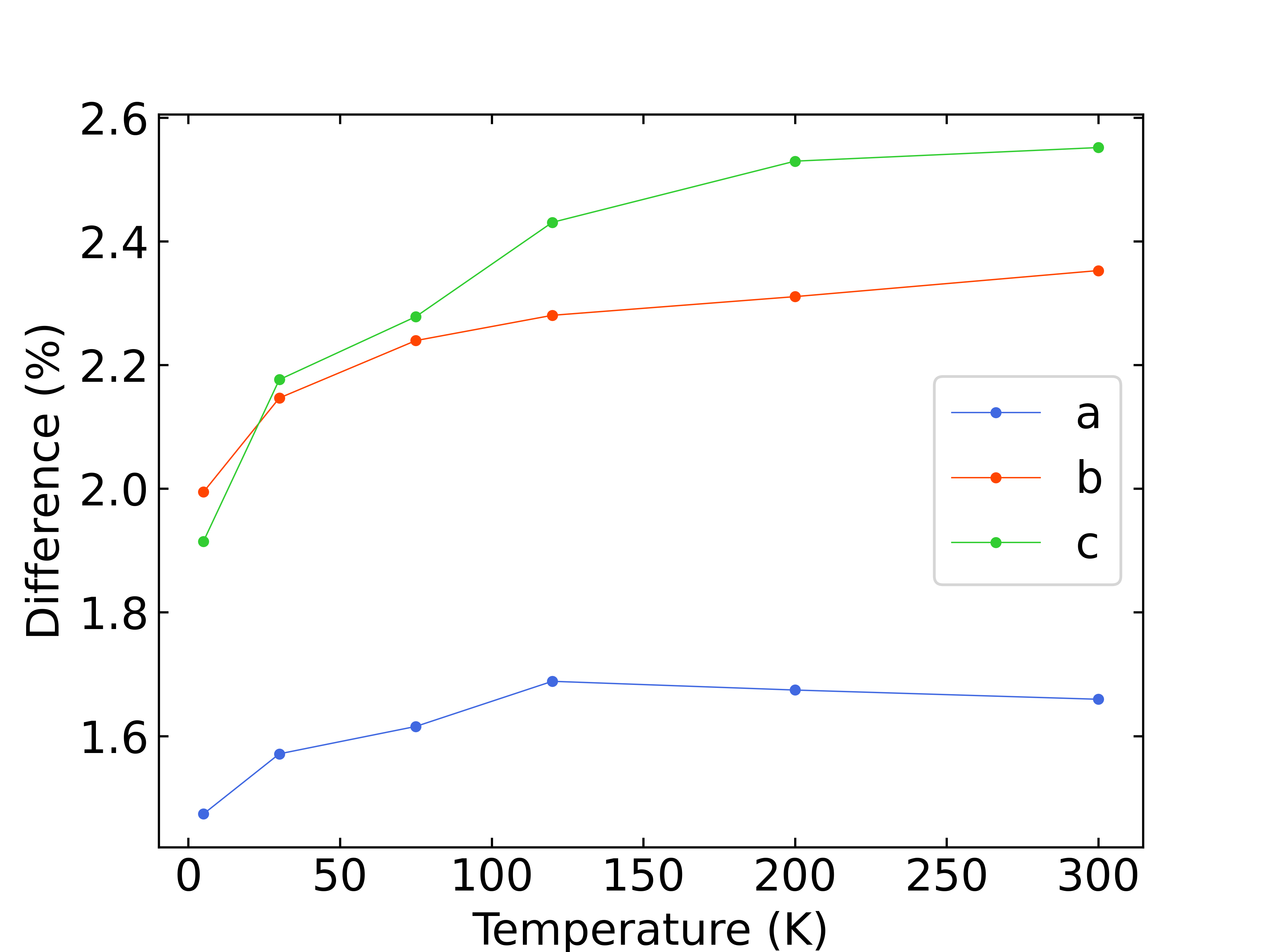}
\caption{Percentage difference between calculated and experimental lattice parameters at different temperatures. Experimental values taken from \cite{yamada2000evaluation} (for a and b) and \cite{capponi1987structure} (for c).  }
\label{thermexp}
\end{figure}

Our thermal expansion data shows satisfactory agreement with the experimentally reported values. We calculate an averaged thermal expansion of $20.36\times10^{-6}$ $K^{-1}$, compared with the values from Yamada \textit{et al.} of $14\times 10^{-6}K^{-1}$ \cite{yamada2000evaluation} (twinned YBCO film) and $13.4\times10^{-6} K^{-1}$ \cite{kawashima1998critical}(for untwinned YBCO film). We obtain linear expansion coefficients $13.4\times10^{-6}$ $K^{-1}$, $13.2\times10^{-6}$ $K^{-1}$, and $34.5\times10^{-6}$ $K^{-1}$ for the a, b and c directions, respectively. Experimental evidence suggests there is a large difference in thermal expansion between the c axis and a and b axes \cite{meingast1991large}, however, the anisotropy in expansion is not as 
pronounced as our DFT results suggest. The discrepancies here may arise due to an inadequate description of the strongly correlated electrons in copper. Experimental data shows\cite{meingast1991large} large, anisotropic jumps at $T_c$ in the a and b axis expansion coefficients due to the superconductivity coupling strongly to the orthorhombic strain. DFT does not describe the superconducting transition, therefore, it is no surprise this is not observed in our data. Furthermore, Fig. \ref{thermexp} shows that past $T_c$ agreement between experiment and our calculated values improves (at least for the rate of change of lattice parameters with respect to temperature). Indeed, the inability of the DFT model to include the superconducting phase will affect our calculated averaged thermal expansion coefficients. We note that there are very few studies that explore the thermal expansion for stoichiometric YBa$_2$Cu$_3$O$_7$. Those that do investigate this show a significant contribution to thermal expansion coefficients based upon oxygen ordering effects in the lattice, an effect not apparent from the fully oxygenated state \cite{nagel2000anomalously}. It is also worth noting that the expansion coefficients are highly dependent on the technique used to grow the sample. For instance, Zeisberger \textit{et al.} report up to 30 \% difference between their thermal expansivities (for melt textured YBCO) and single, untwinned crystals \cite{zeisberger2005measurement}. 

\subsection{Defect Formation Energies}

Table \ref{defE} shows our calculated defect formation energies, in comparison to the literature. \\

\begin{table}[htbp]
\centering

    \centering
    \caption{Comparison of our DFT values for the vacancy defect formation energies of different vacancies in YBCO compared to literature values \cite{liu2023first}$^a$, \cite{murphy2020point}$^b$.}
  
    \begin{tabular}{ccc}
    \hline
   & \multicolumn{1}{c}{Defect formation energy (eV)} \\ \cline{1-3}

        Vacancy type                  & DFT     & Literature DFT  \\ \hline\hline
    V$_{O1}$                 &    0.94       & 0.83$^a$        \\ 
     V$_{O2}$                   &      1.61     & 1.75$^a$   \\ 
     V$_{O3}$                   &      1.57     &    1.74$^a$  \\ 
     V$_{O4}$                    &      1.27     &   1.36$^a$  \\ 
     V$_{Cu1}$    & 2.49 & 2.15$^b$  \\
     V$_{Cu2}$    & 1.84 & 1.76$^b$  \\
     V$_Y$    & 10.85 & 10.82$^b$  \\
     V$_{Ba}$  & 6.87 & 7.17$^b$  \\
     
    \end{tabular}
    \label{defE}
\end{table}

Experiment suggests that that the O1 atom in the Cu-O chains is the most mobile, therefore we would expect it to have the lowest defect formation energy. This is indeed what is observed. The ordering of the oxygen formation energies is also in good agreement with a previous DFT study by Liu \textit{et al.} \cite{liu2023first}. The formation energies for the cation vacancies also follow the trends observed in \cite{murphy2020point}. \\

\subsection{Electronic structure}

\begin{figure}[h!]
\centering
\begin{subfigure}{\linewidth}

\includegraphics[width = \linewidth]{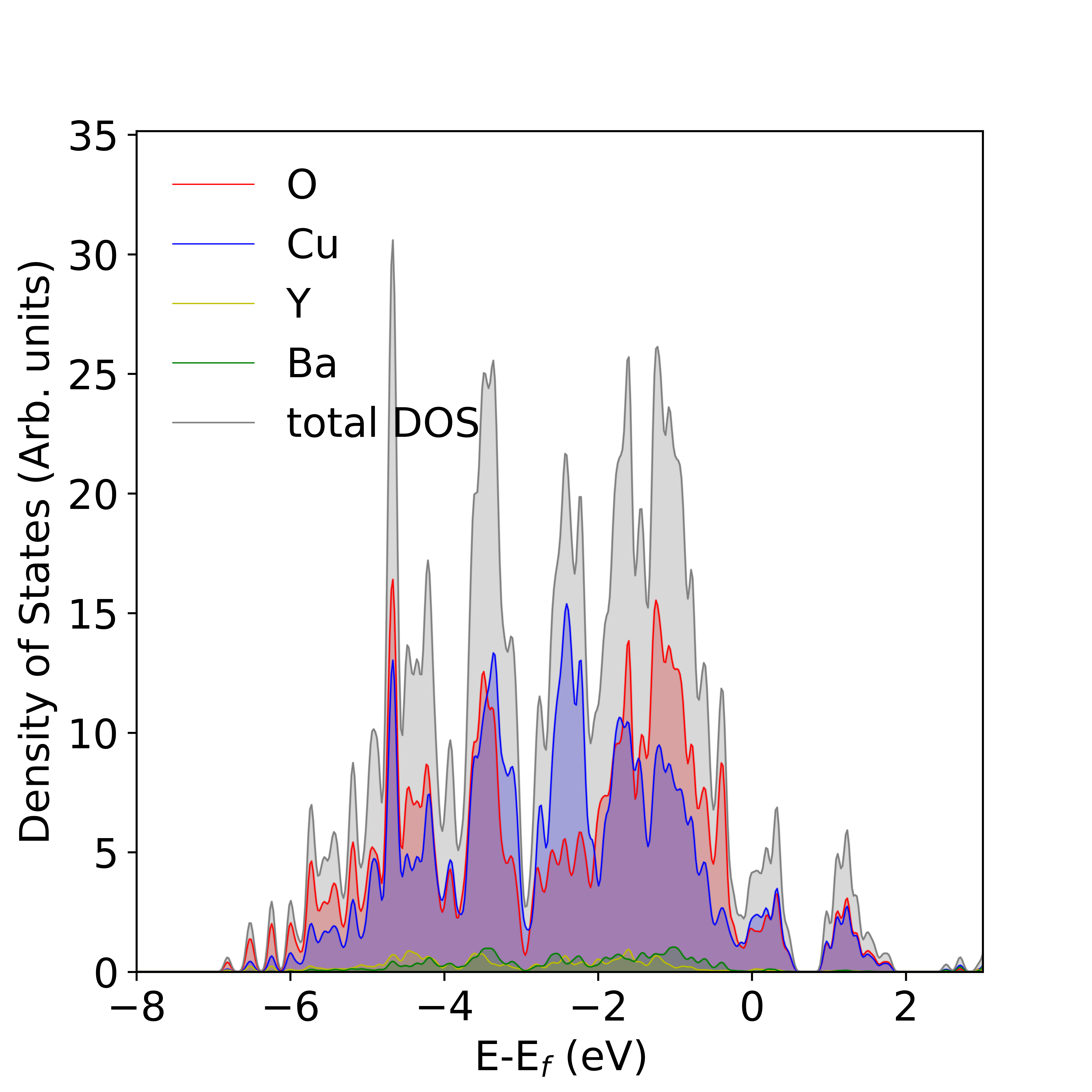}
\caption{}
\label{DOS}
\end{subfigure}
\centering
\begin{subfigure}{\linewidth}

\includegraphics[width = \linewidth]{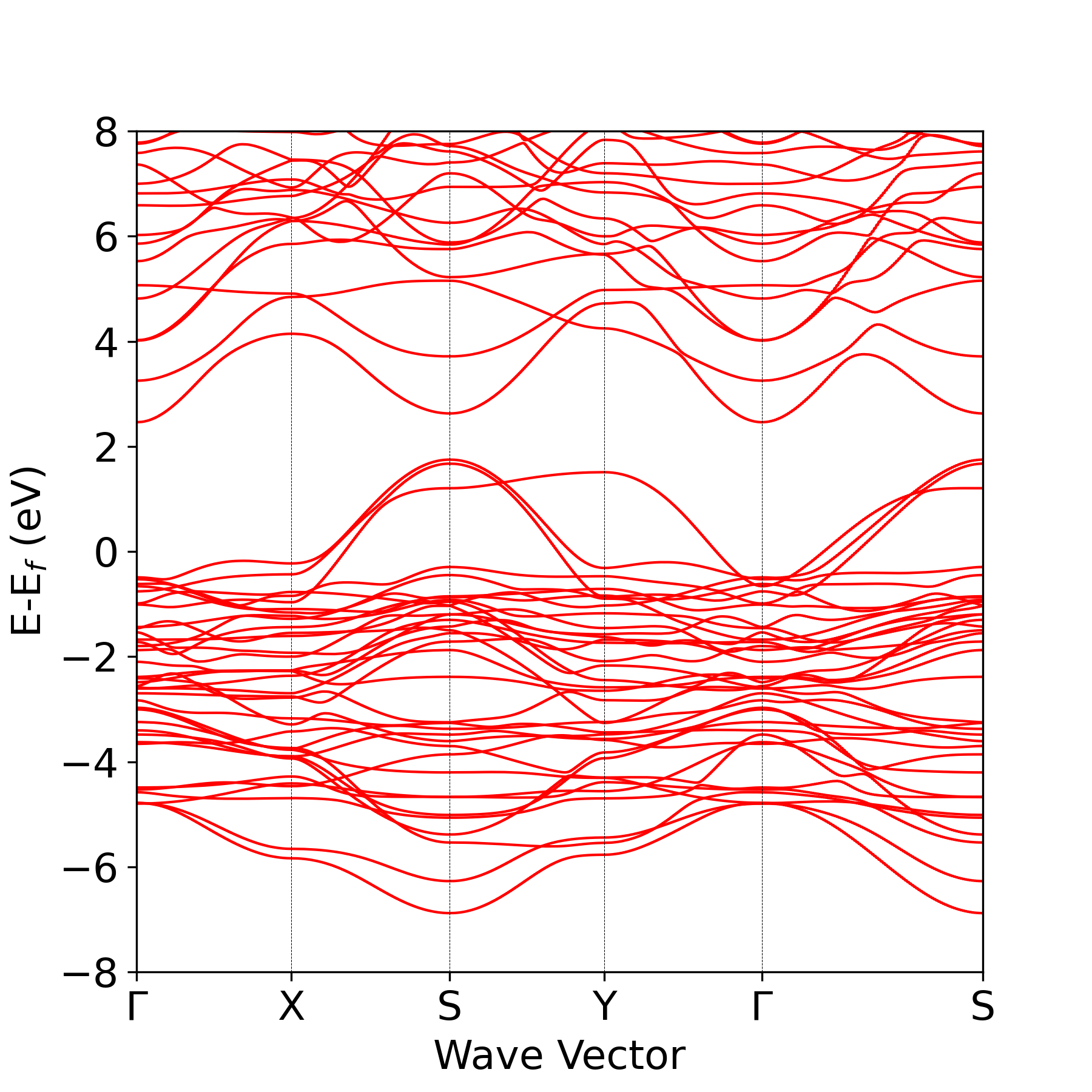}
\caption{}
\label{BS}
\end{subfigure}
\caption{(a) Total and partial DOS overlaid for YBCO. (b) Calculated Band structure of YBCO. Zero energy is taken at the Fermi level. Both normalised such that the Fermi energy is at the 0 value.}
\end{figure}

The electronic Density Of States (DOS) for our DFT study of YBCO is shown in Fig. \ref{DOS}. Our results agree well with other work that utilises local and semi-local exchange functionals \cite{massidda1987electronic,murphy2020point} . Interestingly, there is little difference in the resulting electronic structure from these two levels of theory. We note a small difference between the DOS in our work as compared to other DFT studies. Small peaks just above the Fermi level are observed that are not present in other simulations. These are predominately attributed to oxygen 2p and copper 3d states, and may be due to the Gaussian representation of the wavefunction employed in CP2K, as previous work employed a plane wave basis.\\

The observed band structure (Fig. \ref{BS}) also agrees well with the literature. Near the Fermi level, the material  demonstrates semiconductor-like properties, with a pseudogap below 1 eV. The presence of the valence bands above the Fermi level aligns with the presumption that YBCO is a hole carrier conductor \cite{chu2015hole}. The three bands that cross $E_F$ are attributed to the Cu-O chains and Cu-O$_2$ plane \cite{cheong2024first}. \\

\section{Determination of thermal spike}\label{thermspike}

In order that our TDE simulations are computationally tractable for DFT, we determine first the highest expected PKA energy for our TDE simulations, then determine the length of the thermal spike at this temperature. This allows us to define a minimum cutoff time for our simulations. \\

\begin{figure}[h!]
\centering
    \includegraphics[width=\linewidth]{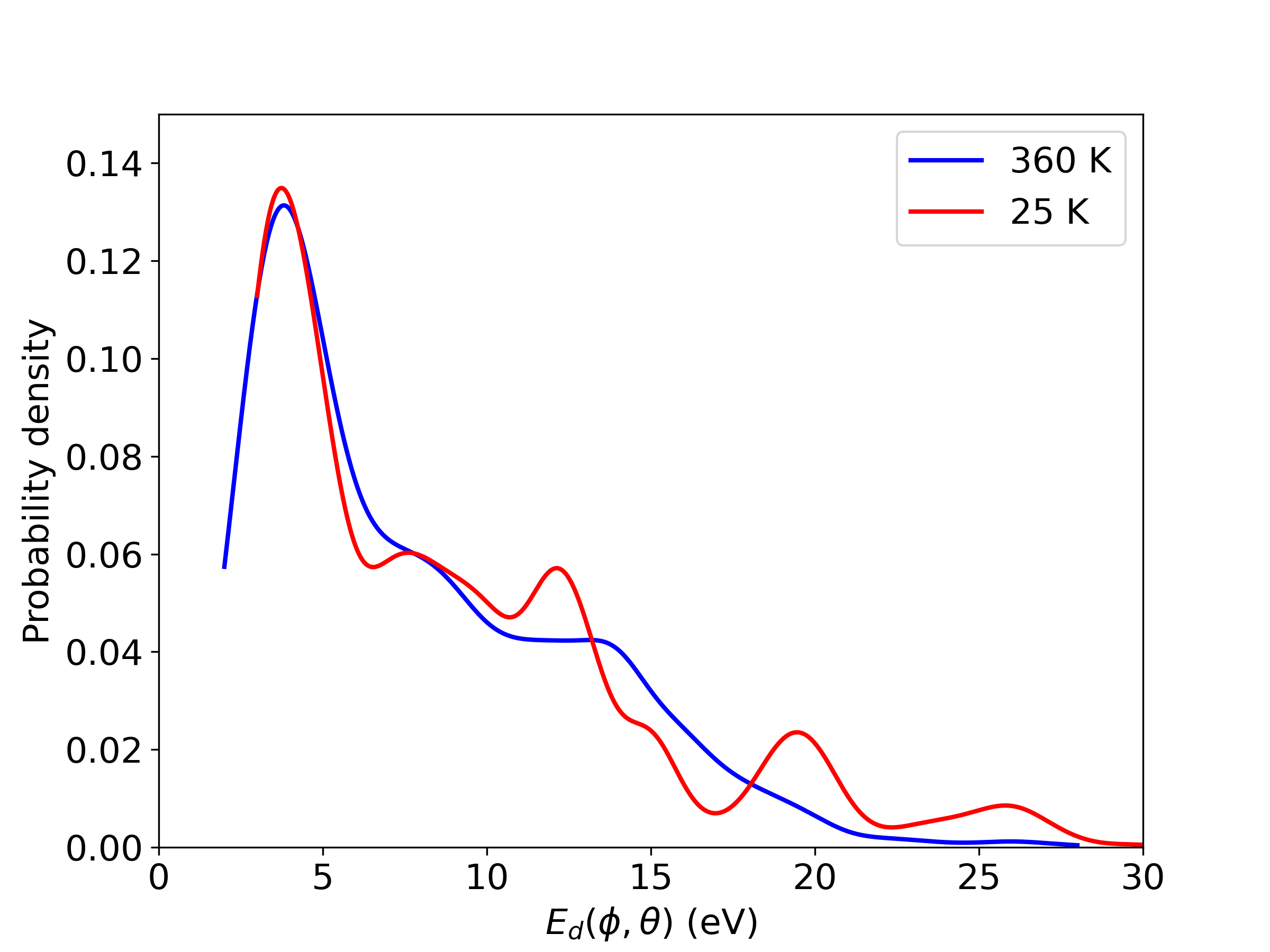}
\caption{Probability distribution function for $E_d(\phi, \theta)$ (for random $\theta$ and $\phi$) in YBCO over 500 directions at 360 K  and 25 K. }
\label{pdistEV}
\end{figure}

Previous TDE results indicate an $E_d$ for O1 of between 9 and 20 eV \cite{gray2022molecular, chaplot1990interatomic, cui1992preferential}. To confirm this range, we performed 500 randomly directed collision cascades for an O1 PKA along vectors within the unit hemisphere oriented with the pole facing the \textit{a}-axis of YBCO. All simulations were performed with the same starting configuration. The probability distribution of threshold displacement energies is shown in Fig. \ref{pdistEV}, where it is evident that the largest displacement energy with a significant probability is $\sim$30 eV. Our TDE simulations follow the same steps outlined in the methodology section. \\

\begin{figure}[h!]
\begin{subfigure}{\linewidth}
\centering
\includegraphics[width=\linewidth]{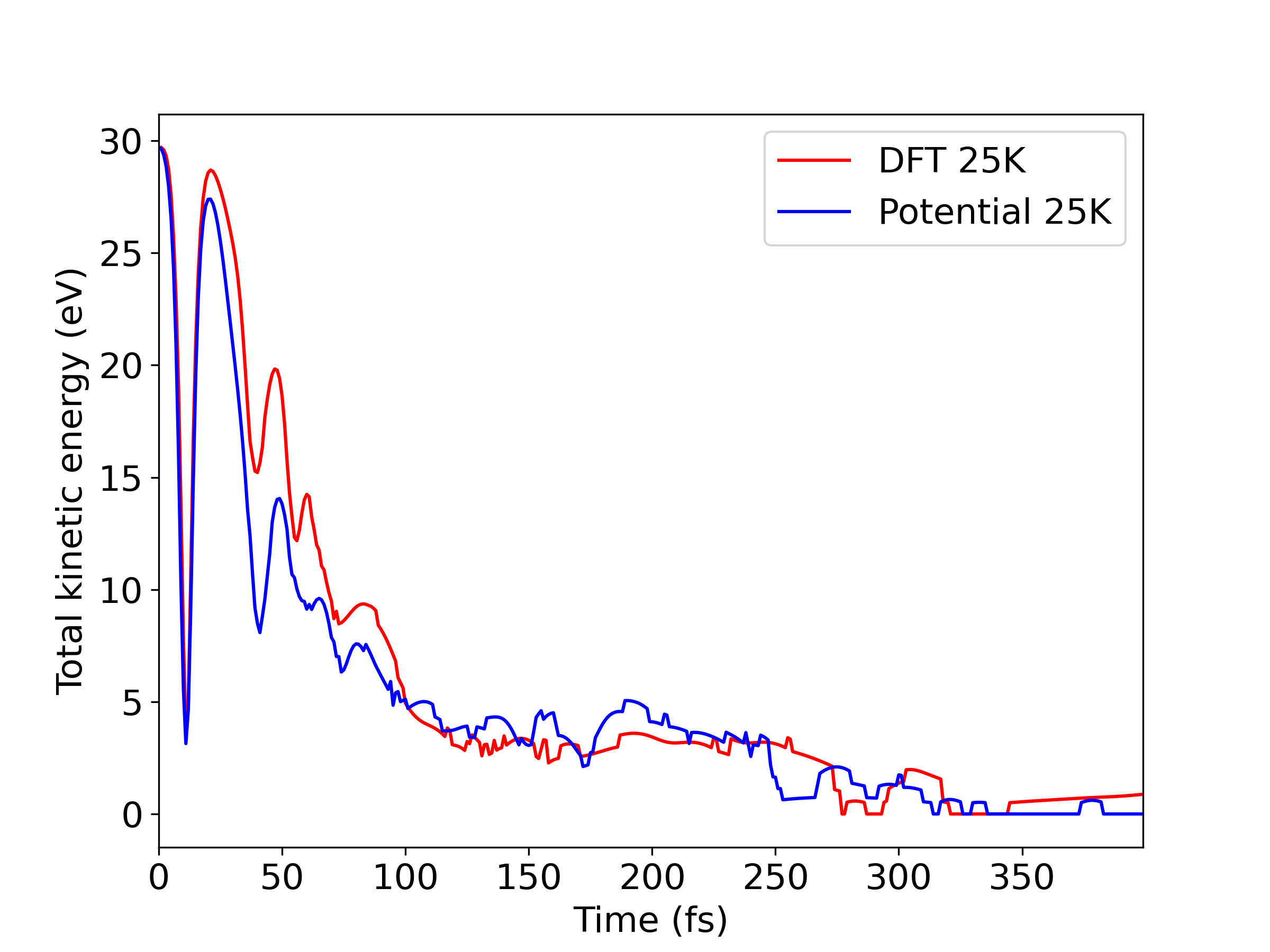}
\caption{25 K}
\end{subfigure}
\begin{subfigure}{\linewidth}
\centering
    \includegraphics[width=\linewidth]{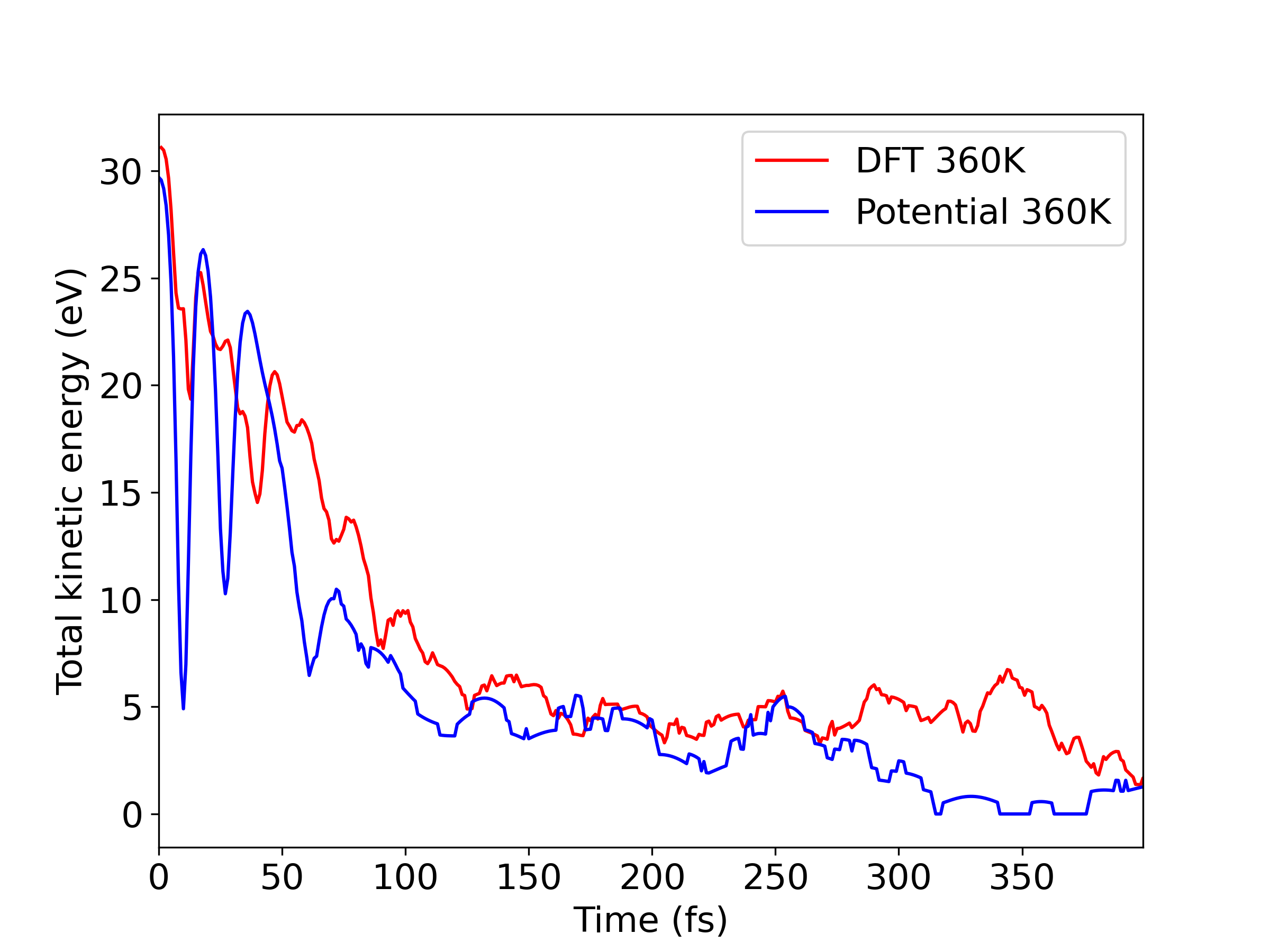}
\caption{360 K}
\end{subfigure}
\caption{Summed kinetic energy for all atoms with kinetic energy above 0.5 eV for a 30 eV cascade.}
\label{SPIKE}
\end{figure}

Therefore, we performed a 30 eV cascade in DFT in a random direction to determine the length of the thermal spike (we adopt this methodology from \cite{dacus2019calculation}). Figures \ref{SPIKE}a) and b) show the thermal spike for both DFT and classical MD simulations, using the same vector direction and energy, at 25 and 360 K. Our results indicate that the thermal energy spike is concluded by around 400 fs. We note that the EP exhibits higher thermal conductivity, and dissipates the heat spike quicker than the DFT model. Nevertheless, for consistency, we keep the defect check time the same between the two models.

% The \nocite command causes all entries in a bibliography to be printed out
% whether or not they are actually referenced in the text. This is appropriate
% for the sample file to show the different styles of references, but authors
% most likely will not want to use it.
%\nocite{*}

%apsrev4-2.bst 2019-01-14 (MD) hand-edited version of apsrev4-1.bst
%Control: key (0)
%Control: author (8) initials jnrlst
%Control: editor formatted (1) identically to author
%Control: production of article title (0) allowed
%Control: page (0) single
%Control: year (1) truncated
%Control: production of eprint (0) enabled
%

\end{document}